\documentclass[letterpaper,11pt]{article}
\usepackage{jheppub} % for details on the use of the package, please
\usepackage{graphicx}
\usepackage{epstopdf}
\usepackage{amsmath, amssymb, float, booktabs}
\usepackage{longtable}
\usepackage{array}
\usepackage{comment}
\usepackage{tikz}
\usepackage{bm}
\usepackage{bbold}
\usepackage{subcaption}
\usepackage[utf8]{inputenc}
\usepackage[all]{xy}
\usepackage{ytableau}
\ytableausetup{smalltableaux, aligntableaux = center}

	\def\CI{{\cal I}}

	\def\CN{{\cal N}}

	\def\CW{{\cal W}}

	\def\a{\alpha}

	\def\k{\kappa}
	
	\def\m{\mu}
	
	\def\ch{\chi}
	\def\G{\Gamma}
	\def\D{\Delta}

	\def\half{\frac{1}{2}}

	\def\vev#1{\langle #1 \rangle}

	\def\tr{{\rm Tr}}

	\def\ft{\mathfrak{t}}

\usepackage{pdflscape}

\title{Dual $\mathcal{N}=1$ Lagrangians for Argyres-Douglas theories}

\author[1, 2]{Minseok Cho}
\author[1]{and Jaewon Song}
\affiliation[1]{Department of Physics, Korea Advanced Institute of Science and Technology\\ 291 Daehak-ro, Yuseong-gu, Daejeon 34141, Republic of Korea}
\affiliation[2]{School of Physics, Korea Institute for Advanced Study\\
85 Hoegiro, Dongdaemun-gu, Seoul, 02455, Republic of Korea}
\emailAdd{minseokcho@kias.re.kr}
\emailAdd{jaewon.song@kaist.ac.kr}

\preprint{KIAS-P26046}
\abstract
{
We find a new set of four-dimensional $\mathcal{N}=1$ Lagrangian gauge theories that flow in the infrared to $\mathcal{N}=2$ Argyres-Douglas theories of type $(A_1, A_{2n})$, $(A_1, A_{2n+1})$, $(A_1, D_{2n+1})$, and $(A_1, D_{2n+2})$. The new dual theories are given by $SU(2)^n$ quiver gauge theories with a number of gauge-singlet flip fields. This description provides dual theories for the previous $\mathcal{N}=1$ Lagrangian realizations given by $Sp(n)$ or $SU(n+1)$ gauge theories, exchanging the rank and the number of nodes in a quiver diagram.
}

\begin{document} 
\maketitle

%%%%%%%%%%%%%%%%%%%%%%%%%%%%%%%%%%%%%%%%%%
%%%%%%%%%%%%%%%%%%%%%%%%%%%%%%%%%%%%%%%%%%
\section{Introduction} \label{sec:intro}

The Argyres-Douglas (AD) theories \cite{Argyres:1995jj, Argyres:1995xn} are interacting $\CN=2$ SCFTs characterized by having Coulomb branch operators of fractional dimensions, thereby admitting no Lagrangians with manifest $\CN=2$ supersymmetry. They are prototypical examples of so-called non-Lagrangian field theories. 
They can nevertheless be reached from Lagrangian field theories as a low-energy limit. These theories were originally obtained by taking a certain scaling limit of the effective field theory on the Coulomb branch of $\CN=2$ supersymmetric Yang-Mills (or QCD) theory, where mutually non-local particles become simultaneously massless. This description is useful for investigating the Coulomb phase of the underlying superconformal field theory (SCFT), but less effective for understanding its conformal phase and spectrum. 

Later, it was found that there are $\CN=1$ gauge theories whose IR fixed points are described by the AD theories with enhanced $\CN=2$ supersymmetry \cite{Maruyoshi:2016tqk, Maruyoshi:2016aim, Agarwal:2016pjo}. This description can be obtained by certain $\CN=1$ -- preserving deformations of $\CN=2$ SCFTs. In particular, it was found that $(A_1, A_N)$ and $(A_1, D_N)$ theories are obtained via $SU$ and $Sp$ gauge theories. 
This construction has been further generalized to a wider class of Argyres-Douglas theories \cite{Xie:2012hs, Wang:2015mra}, which are realized as fixed points of quiver gauge theories \cite{Agarwal:2017roi,Benvenuti:2017bpg}.\footnote{See also \cite{Gadde:2015xta, Agarwal:2018ejn, Gang:2018huc, Razamat:2019vfd, Zafrir:2019hps, Razamat:2020gcc, Zafrir:2020epd, Kang:2023dsa, Maruyoshi:2023mnv, Gang:2023rei} for other examples of SUSY enhancements involving non-Lagrangian SCFTs.}

In this paper, we present alternative Lagrangian gauge theories for the AD theories of type $(A_1, A_N)$ and $(A_1, D_N)$. They are linear quiver gauge theories, whose gauge group is given by $SU(2)_1\times\cdots\times SU(2)_n$. The matter content is given by adjoint chiral multiplets at each gauge node, single bifundamentals linking the neighboring nodes, with an additional pair of fundamentals at one or both ends of the quiver. In addition, we have a number of gauge-singlet flip fields for either `mesons' or `Casimir operators'.
Four distinct choices of the fundamentals at the ends of the quiver (and their respective superpotential) correspond to the  flow to AD theories of type
\begin{align}
(A_1,D_{2n+1})\,,\qquad (A_1,A_{2n})\,,\qquad (A_1,D_{2n+2})\,,\qquad (A_1,A_{2n+1})\,.
\end{align}
We list the four quivers and their respective central charges in Table~\ref{tab:summary}.

\begin{table}[t]
\centering
\small
\setlength{\tabcolsep}{5pt}
\begin{tabular}{ccccc}
\hline
section & quiver & superpotential & flows to & $(a,c)$\\
\hline
\ref{sec:D-odd} & single-ended & \eqref{eq:WD} & $(A_1,D_{2n+1})$ & $\left(\dfrac{n(8n+3)}{8(2n+1)},\ \dfrac n2\right)$\\[8pt]
\ref{sec:A-even} & single-ended & \eqref{eq:WA1} & $(A_1,A_{2n})$ & $\left(\dfrac{n(24n+19)}{24(2n+3)},\ \dfrac{n(6n+5)}{6(2n+3)}\right)$\\[8pt]
\ref{sec:D-even} & double-ended & \eqref{eq:WDA2} & $(A_1,D_{2n+2})$ & $\left(\dfrac n2+\dfrac1{12},\ \dfrac n2+\dfrac16\right)$\\[8pt]
\ref{sec:A-odd} & double-ended & \eqref{eq:WA1A2} & $(A_1,A_{2n+1})$ & $\left(\dfrac{12n^2+19n+2}{24(n+2)},\ \dfrac{3n^2+5n+1}{6(n+2)}\right)$\\[4pt]
\hline
\end{tabular}
\caption{The four theories, defined in Sections~\ref{sec:D-odd}--\ref{sec:A-odd}, and their infrared central charges, which equal the Argyres-Douglas values~\cite{Aharony:2007dj,Shapere:2008zf, Shapere:2008un, Xie:2012hs} for every $n\ge1$.}
\label{tab:summary}
\end{table}

The quiver gauge theory we describe in the current paper can be thought of as a dual gauge theory for the $SU(n+1)$ or $Sp(n)$ gauge theory that flows to $(A_1, D_N)$ or $(A_1, A_N)$ with $N=2n$ or $N=2n+1$ found in \cite{Maruyoshi:2016tqk, Maruyoshi:2016aim, Agarwal:2016pjo}. 
This duality yields integral identities involving elliptic gamma functions. We compute the superconformal index for a few values of $n$ in each family and find that they agree to leading order as power series. In particular, we find that the Schur index reproduces the vacuum character of the associated vertex operator algebra \cite{Beem:2013sza, Buican:2015ina, Cordova:2015nma, Song:2015wta, Song:2017oew}. 

The organization of this paper is as follows: 
In Sections \ref{sec:D-odd A-even} and \ref{sec:D-even A-odd}, we describe the quiver gauge theories in detail and study properties of their IR fixed points and find agreements with that of $(A_1,D_{2n+1})$, $(A_1,A_{2n})$, $(A_1,D_{2n+2})$, and $(A_1,A_{2n+1})$ respectively. 
Then we conclude with possible future directions. 
In Appendix~\ref{sec:methods}, we review the general methods and fix the conventions for the $a$-maximization and the superconformal index.

\section{From \texorpdfstring{$SU(2)^n$}{} quiver to \texorpdfstring{$(A_1,D_{2n+1})$ and $(A_1,A_{2n})$}{(A1,D2n+1)}}
\label{sec:D-odd A-even}

\subsection{\texorpdfstring{$(A_1,D_{2n+1})$}{(A1,D2n+1)} theory}
\label{sec:D-odd}
\begin{figure}[t]
\centering
\begin{tikzpicture}[gauge/.style={circle,draw,minimum size=7.5mm,inner sep=1pt},flavor/.style={rectangle,draw,minimum size=6.5mm,inner sep=3pt}]
\node[gauge] (g1) at (0,0) {$2$};
\node[gauge] (g2) at (1.9,0) {$2$};
\node (dd) at (3.8,0) {$\cdots$};
\node[gauge] (gm) at (5.7,0) {$2$};
\node[gauge] (gn) at (7.6,0) {$2$};
\node[flavor] (q0) at (-2.2,0) {$2$};
\draw (q0)-- node[above]{\footnotesize $q_1$} node[pos=0.5]{$\times$} node[below]{\footnotesize $M_1$} (g1);
\draw (g1)-- node[above]{\footnotesize $q_2$} node[pos=0.5]{$\times$} node[below]{\footnotesize $M_2$} (g2);
\draw (g2)-- node[above]{\footnotesize $q_3$} (dd);
\draw (dd)-- node[above]{\footnotesize $q_{n-1}$} (gm);
\draw (gm)-- node[above]{\footnotesize $q_n$} (gn);
\draw (g1) to[out=115,in=65,looseness=8] node[pos=0.5,above=3pt]{\footnotesize $\phi_1$} (g1);
\draw (g2) to[out=115,in=65,looseness=8] node[pos=0.5,above=3pt]{\footnotesize $\phi_2$} (g2);
\draw (gm) to[out=115,in=65,looseness=8] node[pos=0.5]{$\times$} node[pos=0.5,above=3pt]{\footnotesize $\phi_{n-1}$} node[pos=0.5,right=4pt]{\footnotesize $X_{n-1}$} (gm);
\draw (gn) to[out=115,in=65,looseness=8] node[pos=0.5]{$\times$} node[pos=0.5,above=3pt]{\footnotesize $\phi_n$} node[pos=0.5,right=4pt]{\footnotesize $X_n$} (gn);
\end{tikzpicture}
\caption{The quiver diagram for the $\CN=1$ theory flowing to $(A_1, D_{2n+1})$ theory. As usual, the circular nodes correspond to the gauge nodes, with their adjoints $\phi_i$ drawn as loops; the rectangular node carries the flavor symmetry $SU(2)_0$ rotating the pair of fundamentals $q_1$.  The cross on a line denotes a flip field: $M_i$ ($i=1, \cdots, \lceil n/2\rceil$) is a flip field for $q_i q_i$, and $X_k$ ($k = \lfloor n/2\rfloor+1, \cdots, n$) is a flip field for $\tr \phi_k^2$.}
\label{fig:D-odd}
\end{figure}
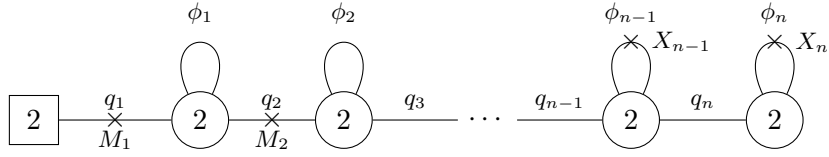

Consider a quiver gauge theory with gauge group  $SU(2)_1\times\cdots\times SU(2)_n$, with an adjoint chiral multiplet $\phi_i$ at each gauge node and a bifundamental $q_i$ of $SU(2)_{i-1}\times SU(2)_i$, $1\le i\le n$. Here $SU(2)_0$ is a flavor symmetry. 
%We find that upon certain (flip) deformations, the theory flows to $(A_1,D_{2n+1})$ and $(A_1,A_{2n})$ Argyres-Douglas theories.
In addition, introduce a set of gauge singlets $M_i$, $X_k$ with the superpotential coupling given as
\begin{align}\label{eq:WD}
W&=\sum_{i=1}^{n-1}\phi_i q_{i+1} \phi_{i+1} q_{i+1}
+\sum_{i=1}^{\lceil n/2\rceil}M_i q_i q_i+\sum_{k=\lfloor n/2\rfloor+1}^{n}X_k\tr\phi_k^2\,.
\end{align}
See Figure~\ref{fig:D-odd} for a graphical presentation of our quiver gauge theory and Table~\ref{tab:D-odd} for the list of matter fields and their charges. Notice that unlike usual quivers, we have exactly \emph{one} bifundamental (rather than a pair) for each edge. This is free of Witten's $SU(2)$ anomaly and allowed \cite{Witten:1982fp}. 
As an $\CN=1$ gauge theory, it has $SU(2) \times U(1)_f$ flavor symmetry, where the first factor comes from rotating a pair of fundamentals $q_1$, and the $U(1)_f$ is the unique anomaly-free abelian symmetry besides $R$.

\begin{table}[t]
\centering
\renewcommand{\arraystretch}{1.6}
\setlength{\tabcolsep}{5pt}
\begin{tabular}{llcc}
\hline
field & representation & $R$ & $f$\\
\hline
$\phi_i$, $1\le i\le n$ & $\mathbf 3$ of $SU(2)_i$ & $\dfrac{2(2n+1-2i)}{3(2n+1)}$ & $\dfrac{2(2n+1-2i)}{2n+1}$\\
$q_i$, $1\le i\le n$ & $(\mathbf 2_{i-1},\mathbf 2_{i})$ & $1-\dfrac{4(n+1-i)}{3(2n+1)}$ & $-\dfrac{4(n+1-i)}{2n+1}$\\
$M_i$, $1\le i\le\lceil\tfrac n2\rceil$ & $\mathbf 1$ & $\dfrac{8(n+1-i)}{3(2n+1)}$ & $\dfrac{8(n+1-i)}{2n+1}$\\
$X_k$, $\lfloor\tfrac n2\rfloor+1\le k\le n$ & $\mathbf 1$ & $2-\dfrac{4(2n+1-2k)}{3(2n+1)}$ & $-\dfrac{4(2n+1-2k)}{2n+1}$\\
\hline
\end{tabular}
\caption{Matter content of the $\CN=1$ theory that flows to $(A_1,D_{2n+1})$: gauge representation, infrared $R$-charge, and $U(1)_f$ charge of each field. Here the subscript $i$ denotes the $i$-th gauge factor, except for $i=0$, which is the flavor symmetry $SU(2)_0$.}
\label{tab:D-odd}
\end{table}

In order for the theory to flow to a superconformal fixed point, we should have non-anomalous $R$-symmetry. Therefore, we demand the anomaly-free condition $\tr R GG = 0$ for each gauge node, which gives $n$ constraints. In addition, each term in the superpotential \eqref{eq:WD} imposes additional constraints. Combining these conditions, we have
\begin{align}
\begin{split}
2+2(R_{\phi_i}-1)+(R_{q_i}-1)+(R_{q_{i+1}}-1)&=0\,,\qquad 1\le i\le n-1\,,\\
2+2(R_{\phi_n}-1)+(R_{q_n}-1)&=0\,,\label{eq:D-nodes}\\
R_{\phi_i}+2R_{q_{i+1}}+R_{\phi_{i+1}}=2\,,\qquad R_{M_i}+2R_{q_i}&=2\,,\qquad R_{X_k}+2R_{\phi_k}=2\,,
\end{split}
\end{align}
where $R_{\text{field}}$ denotes the $R$-charge for the superfield. 
This condition leaves one free parameter, which we have to fix using $a$-maximization \cite{Intriligator:2003jj}. Writing $x\equiv R_{\phi_n}$, all the $R$-charges can be written as:
\begin{align}
\begin{aligned}
R_{\phi_i}&=(2n+1-2i)x\,,& R_{q_i}&=1-2(n+1-i)x\,,\\
R_{M_i}&=4(n+1-i)x\,,& R_{X_k}&=2-2(2n+1-2k)x\,.
\end{aligned}
\label{eq:D-charges}
\end{align}
Then, we compute the anomaly coefficients for the $R$-symmetry to obtain
\begin{align}
\tr R^3=3nx\left[n(2n+1)^2x^2-(4n+1)(2n+1)x+(4n+1)\right]\,, \quad\tr R=-3nx\,.
\end{align}
Now, using the relations between central charges and the trace anomalies \cite{Anselmi:1997am}
\begin{align}
    a = \frac{3}{32} \left( 3 \tr R^3 - \tr R \right) \ , \quad c = \frac{1}{32} (9 \tr R^3 - 5 \tr R) \ ,
\end{align}
we get
\begin{align}
a(x)&=\frac{9nx}{32}\left[3n(2n+1)^2x^2-3(4n+1)(2n+1)x+12n+4\right]\,.
\label{eq:D-a}
\end{align}
The superconformal $R$-symmetry should locally maximize the central charge $a(x)$. The local maximum is given by $x=\tfrac{2}{3(2n+1)}$, giving the superconformal $R$-symmetry at the IR fixed point. We list the $R$-charges of the elementary fields in Table~\ref{tab:D-odd}. 
The resulting central charges are given as
\begin{align}
(a,c)=\left(\frac{n(8n+3)}{8(2n+1)},\ \frac n2\right)\,,
\end{align}
that precisely agree with the values of $(A_1,D_{2n+1})$ \cite{Aharony:2007dj, Shapere:2008un}.

Let us look at the spectrum of this theory. 
It turns out that the Coulomb branch operators of the $(A_1,D_{2n+1})$ theory are realized by $M_i$ with $1\le i\le\lceil n/2\rceil$ and the $\tr\phi_k^2$ with $1\le k\le\lfloor n/2\rfloor$, which combine to give an $n$-dimensional Coulomb branch. Their scaling dimensions can be obtained by $\Delta = \frac{3}{2} R$, which are $2-\tfrac{2(2i-1)}{2n+1}$ for $M_i$ and $2-\tfrac{4k}{2n+1}$ for $\tr\phi_k^2$. The $M_i$ give $2-\tfrac{2m}{2n+1}$ for every odd $m\le n$ and the $\tr\phi_k^2$ for every even $m\le n$, so these $n$ operators have dimensions $\{2-\tfrac{2m}{2n+1}\}_{m=1}^{n}$, the Coulomb-branch spectrum of $(A_1,D_{2n+1})$. The length of the quiver equals the rank of the Argyres-Douglas theory.

The theory \eqref{eq:WD} has the $SU(2)$ flavor symmetry $SU(2)_0$, rotating the two fundamentals in $q_1$, which maps to the $SU(2)$ flavor symmetry of the $(A_1,D_{2n+1})$ theory. The flavor central charge $k_F$ is given by the mixed 't Hooft anomaly $k_F \delta^{AB}=-6\tr R T^AT^B$, with $T^A$ being the flavor symmetry generators normalized as $\tr T^AT^B=\tfrac12\delta^{AB}$ in the fundamental representation. We obtain
\begin{align}
    k_{SU(2)}=\frac{8n}{2n+1} \,,
\end{align}
which is precisely the value of the $(A_1,D_{2n+1})$ theory. 
We find the mesons $q_1\phi_1q_1$ form an $SU(2)_0$ triplet with $\Delta=2$ and $f=-2$, corresponding to the moment map for the $SU(2)$ flavor symmetry.

For $n=1$, we obtain a gauge theory with a single $SU(2)$ factor. This theory is identical to the known theory of \cite{Maruyoshi:2016aim} that flows to $(A_1, A_3) = (A_1, D_3)$.\footnote{The description in \cite{Agarwal:2016pjo} gives a slightly different $\CN=1$ Lagrangian for $(A_1, A_3)=(A_1, D_3)$ theory, which we find to be realized in Section \ref{sec:A-odd}.} However, for $n \ge 2$, our description is different from the one given in \cite{Agarwal:2016pjo} for the $(A_1, D_{2n+1})$ theory, which is realized as $Sp(n)$ gauge theory. We claim that these two gauge theories ($Sp(n)$ and $SU(2)^n$) are infrared dual to each other with enhanced supersymmetry at the fixed point. 

Our claim implies that the superconformal indices of the two dual gauge theories coincide. The index of our quiver gauge theory can be written explicitly as a contour integral of elliptic gamma functions as
\begin{align}
\begin{split}
\CI_{\text{quiver}}^{(A_1, D_{2n+1})} &=\frac{\prod_{i=1}^{\lceil n/2\rceil}\Gamma\left(\left(\tfrac{pq}t\right)^{2-\frac{2(2i-1)}{2n+1}}\right)}{\prod_{k=\lfloor n/2\rfloor+1}^{n}\Gamma\left(\left(\tfrac{pq}t\right)^{\frac{2(2n+1-2k)}{2n+1}}\right)}\,\frac{\kappa^n}{2^n}\oint\prod_{i=1}^n\frac{dz_i}{2\pi iz_i}\\
&\quad\times \prod_{i=1}^n\frac{\Gamma\left(z_i^{\pm2,0}\left(\tfrac{pq}t\right)^{\frac{2n+1-2i}{2n+1}}\right)}{\Gamma(z_i^{\pm2})} 
\prod_{i=1}^n\Gamma\left(z_{i-1}^{\pm}z_i^{\pm}\left(\tfrac{pq}t\right)^{\frac{4i-2n-3}{2(2n+1)}}\,t^{\frac12}\right)\,,\label{eq:gammaD}
\end{split}
\end{align}
where $z_0\equiv a$ should be understood as the $SU(2)_0$ flavor fugacity. The symbol $\Gamma(z)$ denotes the elliptic gamma function. See Appendix~\ref{sec:methods} for the details.
The other side of the duality is given by the $Sp(n)$ gauge theory. The index on this dual frame is computed as an integral over the maximal torus of $Sp(n)$~\cite{Agarwal:2016pjo} as
\begin{align}
\begin{split}
&\CI_{Sp(n)}^{(A_1, D_{2n+1})} =\frac{\prod_{m=1}^{n}\Gamma\left(\left(\tfrac{pq}t\right)^{2-\frac{2m}{2n+1}}\right)}{\prod_{i=1}^{n}\Gamma\left(\left(\tfrac{pq}t\right)^{\frac{2i}{2n+1}}\right)}\,\frac{\kappa^n}{2^nn!}\,\oint\prod_{i=1}^n\frac{dz_i}{2\pi iz_i}\prod_{i=1}^n\frac{\Gamma\left(z_i^{\pm2,0}\left(\tfrac{pq}t\right)^{\frac1{2n+1}}\right)}{\Gamma(z_i^{\pm2})}\\
&\quad\times\prod_{1\le i<j\le n}\frac{\Gamma\left(z_i^{\pm}z_j^{\pm}\left(\tfrac{pq}t\right)^{\frac1{2n+1}}\right)}{\Gamma(z_i^{\pm}z_j^{\pm})}
\prod_{i=1}^n\Gamma\left(z_i^{\pm}a^{0,\pm2}\left(\tfrac{pq}t\right)^{\frac{n}{2n+1}}t^{\frac12}\right)\,\Gamma\left(z_i^{\pm}\left(\tfrac{pq}t\right)^{-\frac{n}{2n+1}}t^{\frac12}\right)\,.\label{eq:gammaDsqcd}
\end{split}
\end{align}
The duality implies that the RHS of \eqref{eq:gammaD} equals the RHS of \eqref{eq:gammaDsqcd}. An analytic proof of this identity would be desirable. Instead, we explicitly compute the full superconformal index for $n=2$ and $n=3$ cases as a power series in $\mathfrak{t} \equiv (pq)^{1/6}$, to get (see Appendix \ref{sec:methods} for the precise definitions of symbols)
\begin{align}
\begin{split}
\CI_{(A_1,D_5)}&=1+\ft^{12/5}\xi^{12/5}+\ft^{16/5}\xi^{16/5}-\ft^{17/5}\xi^{2/5}\chi_{2}(y)+3\ft^{4}\xi^{-2}-\ft^{21/5}\xi^{6/5}\chi_{2}(y)\\
&\quad+\ft^{22/5}\xi^{-8/5}+\ft^{24/5}\xi^{24/5}+\ft^{26/5}\xi^{-4/5}+\ft^{27/5}\xi^{12/5}\chi_{2}(y)+\ft^{28/5}\xi^{28/5}\\
&\quad-\ft^{29/5}\xi^{14/5}\chi_{2}(y)-4\ft^{6}+\dots\,,\label{eq:fullD5}
\end{split}
\\
\begin{split}
\CI_{(A_1,D_7)}&=1+\ft^{16/7}\xi^{16/7}+\ft^{20/7}\xi^{20/7}-\ft^{23/7}\xi^{2/7}\chi_{2}(y)+\ft^{24/7}\xi^{24/7}-\ft^{27/7}\xi^{6/7}\chi_{2}(y)\\
&\quad+3\ft^{4}\xi^{-2}+\ft^{30/7}\xi^{-12/7}-\ft^{31/7}\xi^{10/7}\chi_{2}(y)+\ft^{32/7}\xi^{32/7}+\ft^{34/7}\xi^{-8/7}\\
&\quad+\ft^{36/7}\xi^{36/7}+\ft^{37/7}\xi^{16/7}\chi_{2}(y)+\ft^{38/7}\xi^{-4/7}-\ft^{39/7}\xi^{18/7}\chi_{2}(y)+2\ft^{40/7}\xi^{40/7}\\
&\quad+\ft^{41/7}\xi^{20/7}\chi_{2}(y)-4\ft^{6}+\dots\,,\label{eq:fullD7}
\end{split}
\end{align}
where $\chi_m$ denotes the character of the $m$-dimensional representation of $SU(2)$. Here we set the flavor fugacity $a=1$ for simplicity. This index indeed agrees with the one obtained from the deformed $Sp(n)$ SQCD of~\cite{Agarwal:2016pjo}, whose index is computed using a different integral of elliptic gamma functions.
We can take the Schur limit of the index to obtain
\begin{align}
\CI_{(A_1, D_5)} &= 1+3q+9q^2+22q^3+51q^4+105q^5+212q^6+402q^7+744q^8+\dots\,,\label{eq:schurD5}\\
\CI_{(A_1, D_7)} &= 1+3q+9q^2+22q^3+51q^4+108q^5+221q^6+426q^7+801q^8+\dots\,,\label{eq:schurD7}
\end{align}
which agrees with the vacuum characters of $\widehat{\mathfrak{su}}(2)_{-8/5}$ and $\widehat{\mathfrak{su}}(2)_{-12/7}$, respectively. This is consistent with the associated vertex operator algebra (VOA) for $(A_1, D_{2n+1})$ being given by $\mathfrak{su}(2)_{-\frac{4n}{2n+1}}$~\cite{Cordova:2015nma, Buican:2015ina, Song:2015wta}. 
Therefore, we claim our gauge theory flows to the $(A_1, D_{2n+1})$ theory in the infrared, exhibiting supersymmetry enhancement.

\subsection{\texorpdfstring{$(A_1,A_{2n})$}{(A1,A2n)} theory}
\label{sec:A-even}

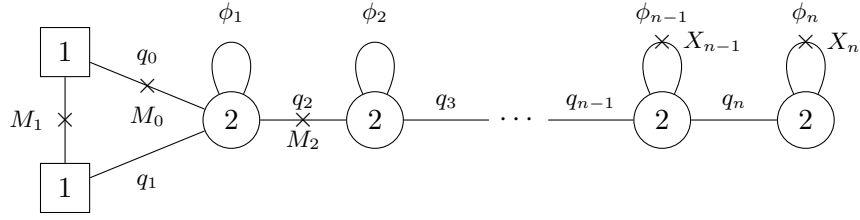
\begin{figure}[t]
\centering
\begin{tikzpicture}[gauge/.style={circle,draw,minimum size=7.5mm,inner sep=1pt},flavor/.style={rectangle,draw,minimum size=6.5mm,inner sep=3pt}]
\node[gauge] (g1) at (0,0) {$2$};
\node[gauge] (g2) at (1.9,0) {$2$};
\node (dd) at (3.8,0) {$\cdots$};
\node[gauge] (gm) at (5.7,0) {$2$};
\node[gauge] (gn) at (7.6,0) {$2$};
\node[flavor] (qa) at (-2.2,0.9) {$1$};
\node[flavor] (qb) at (-2.2,-0.9) {$1$};
\draw (qa)-- node[pos=0.5]{$\times$} node[pos=0.5,above=4pt]{\footnotesize $q_0$} node[pos=0.5,below=4pt]{\footnotesize $M_0$} (g1);
\draw (qb)-- node[pos=0.5,below=4pt]{\footnotesize $q_1$} (g1);
\draw (qa)-- node[pos=0.5]{$\times$} node[pos=0.5,left=4pt]{\footnotesize $M_1$} (qb);
\draw (g1)-- node[above]{\footnotesize $q_2$} node[pos=0.5]{$\times$} node[below]{\footnotesize $M_2$} (g2);
\draw (g2)-- node[above]{\footnotesize $q_3$} (dd);
\draw (dd)-- node[above]{\footnotesize $q_{n-1}$} (gm);
\draw (gm)-- node[above]{\footnotesize $q_n$} (gn);
\draw (g1) to[out=115,in=65,looseness=8] node[pos=0.5,above=3pt]{\footnotesize $\phi_1$} (g1);
\draw (g2) to[out=115,in=65,looseness=8] node[pos=0.5,above=3pt]{\footnotesize $\phi_2$} (g2);
\draw (gm) to[out=115,in=65,looseness=8] node[pos=0.5]{$\times$} node[pos=0.5,above=3pt]{\footnotesize $\phi_{n-1}$} node[pos=0.5,right=4pt]{\footnotesize $X_{n-1}$} (gm);
\draw (gn) to[out=115,in=65,looseness=8] node[pos=0.5]{$\times$} node[pos=0.5,above=3pt]{\footnotesize $\phi_n$} node[pos=0.5,right=4pt]{\footnotesize $X_n$} (gn);
\end{tikzpicture}
\caption{The quiver diagram for the $\CN=1$ theory flowing to $(A_1, A_{2n})$ theory. The cross on a line denotes a flip field: $M_0$ is a flip field for $q_0 \phi q_0$, $M_1$ for $q_0 q_1$, $M_i$ $(i=2, \cdots, \lfloor n/2\rfloor)$ for $q_i q_i$, and $X_k$ $(k = \lceil n/2\rceil, \cdots, n)$ is for $\tr \phi_k^2$.}
\label{fig:A-even}
\end{figure}

Consider the same matter content as the one we discussed for the $(A_1, D_{2n+1})$ theory. Now, consider a slightly different superpotential that breaks the flavor symmetry $SU(2)_0$. We do this by splitting the two fundamentals of $SU(2)_1$ into $q_0$ and $q_1$. We choose the superpotential as
\begin{align}
\begin{split}
    W&=\sum_{i=1}^{n-1}\phi_i q_{i+1} \phi_{i+1} q_{i+1} +q_1\phi_1q_1\\
    &\quad + M_0 q_0\phi_1 q_0 + M_1 q_0q_1+\sum_{i=2}^{\lfloor n/2\rfloor}M_i q_iq_i + \sum_{k=\lceil n/2\rceil}^{n}X_k\tr\phi_k^2 \,,\label{eq:WA1}
\end{split}
\end{align}
and once again $M_i$ are the flip fields for the `mesonic' operators, and $X_k$ are the flip fields for the `Coulomb branch' operators. 
See Figure~\ref{fig:A-even} for a graphical presentation of our quiver gauge theory, and Table~\ref{tab:A-even} for the matter content and the charges.

\begin{table}[h]
\centering
\renewcommand{\arraystretch}{1.6}
\setlength{\tabcolsep}{5pt}
\begin{tabular}{llcc}
\hline
field & representation & $R$ & $f$\\
\hline
$\phi_i$, $1\le i\le n$ & $\mathbf 3$ of $SU(2)_i$ & $\dfrac{2(2n+1-2i)}{3(2n+3)}$ & $\dfrac{2(2n+1-2i)}{2n+3}$\\
$q_0$ & $\mathbf 2$ of $SU(2)_1$ & $\dfrac{8}{3(2n+3)}$ & $-\dfrac{6n+1}{2n+3}$\\
$q_1$ & $\mathbf 2$ of $SU(2)_1$ & $1-\dfrac{2n-1}{3(2n+3)}$ & $-\dfrac{2n-1}{2n+3}$\\
$q_i$, $2\le i\le n$ & $(\mathbf 2_{i-1},\mathbf 2_{i})$ & $1-\dfrac{4(n+1-i)}{3(2n+3)}$ & $-\dfrac{4(n+1-i)}{2n+3}$\\
$M_0$ & $\mathbf 1$ & $\dfrac{4(2n+1)}{3(2n+3)}$ & $\dfrac{4(2n+1)}{2n+3}$\\
$M_1$ & $\mathbf 1$ & $\dfrac{8n}{3(2n+3)}$ & $\dfrac{8n}{2n+3}$\\
$M_i$, $2\le i\le\lfloor\tfrac{n}{2}\rfloor$ & $\mathbf 1$ & $\dfrac{8(n+1-i)}{3(2n+3)}$ & $\dfrac{8(n+1-i)}{2n+3}$\\
$X_k$, $\lceil\tfrac{n}{2}\rceil\le k\le n$ & $\mathbf 1$ & $2-\dfrac{4(2n+1-2k)}{3(2n+3)}$ & $-\dfrac{4(2n+1-2k)}{2n+3}$\\
\hline
\end{tabular}
\caption{Matter content of the $\CN=1$ theory that flows to $(A_1,A_{2n})$: gauge representation, infrared $R$-charge, and $U(1)_f$ charge of every field.}
\label{tab:A-even}
\end{table}

Repeating the same procedure as in Section~\ref{sec:D-odd}, we obtain the $R$-charges as in Table~\ref{tab:A-even}. The resulting central charges are
\begin{align}
(a,c)=\left(\frac{n(24n+19)}{24(2n+3)},\ \frac{n(6n+5)}{6(2n+3)}\right)\,,
\end{align}
which agrees with the values of $(A_1,A_{2n})$ theory \cite{Shapere:2008zf}.
Our quiver theory does not have any flavor symmetry besides $U(1)_f \subset SU(2)_R \times U(1)_r$. This agrees with the expectation since $(A_1, A_{2n})$ theory does not have any flavor symmetry.

The Coulomb branch operators of the $(A_1,A_{2n})$ theory are given by $M_i$ with $i= 0, 1,\allowbreak\cdots, \lfloor n/2\rfloor$, and $\tr\phi_i^2$ with $i= 1, \cdots, \lceil n/2\rceil-1$. Their scaling dimensions are $\tfrac{2(2n+1)}{2n+3}$ for $M_0$, $\tfrac{4(n+1-i)}{2n+3}$ for $M_{i=1, \cdots, \lfloor n/2\rfloor}$ and $\tfrac{2(2n+1-2i)}{2n+3}$ for $\tr\phi_i^2$. Combining them, we have $n$ operators with scaling dimensions $\{\tfrac{2m}{2n+3}\}_{m=n+2}^{2n+1}$, identical to the Coulomb-branch spectrum of $(A_1,A_{2n})$ theory.

We can understand this theory from the RG flow $(A_1,D_{2n+1})\to(A_1,A_{2n})$ triggered by nilpotent Higgsing of the flip field for the moment map~\cite{Maruyoshi:2016aim}. 
The theories described by \eqref{eq:WD} and \eqref{eq:WA1} have the same matter content except for the flip fields. 
We obtain the latter by flipping the moment map operator, and then by nilpotent Higgsing the flip field.  
The operator $q_1^2\phi_1$ is the moment map in the $(A_1, D_{2n+1})$ theory, which has $\Delta=2$ and $f=-2$.
Upon nilpotent Higgsing via $\vev{M} = \sigma^+$, one out of three components survives in the superpotential term \eqref{eq:WA1}. The fluctuation around the vacuum expectation value gives rise to the flip deformation term $M_0q_0^2\phi_1$. Upon removing the Nambu-Goldstone bosons that are eaten, we get the superpotential in \eqref{eq:WA1}. For even $n$, the operator $\tr\phi^2_{n/2}$ hits the unitarity bound along the RG flow and decouples, so \eqref{eq:WA1} contains the additional flip field $X_{n/2}$. For odd $n$, the singlet $M_{(n+1)/2}$ of \eqref{eq:WD} becomes free and decouples.

For $n=1$, the theory is given by a simple $SU(2)$ gauge theory with $\phi_1$, $q_0$, $q_1$, $M_0$, $M_1$ and $X_1$. In this case, $M_1$ violates the unitarity bound $R>2/3$, so it gets decoupled along the RG flow. Upon decoupling $M_1$, we get the description of  \cite{Maruyoshi:2016tqk} (upon removing the decoupled fields there as well) whose central charges are $(a,c)=(\tfrac{43}{120},\tfrac{11}{30})$, identical to that of $H_0=(A_1, A_2)$ theory. 

The superconformal index of this quiver theory takes the integral form,
\begin{align}
\begin{split}
\CI^{(A_1, A_{2n})}_{\text{quiver}}&=\frac{\Gamma\left(\left(\tfrac{pq}t\right)^{\frac{2(2n+1)}{2n+3}}\right)\,\Gamma\left(\left(\tfrac{pq}t\right)^{\frac{4n}{2n+3}}\right)\prod_{i=2}^{\lfloor n/2\rfloor}\Gamma\left(\left(\tfrac{pq}t\right)^{\frac{4(n+1-i)}{2n+3}}\right)}{\prod_{k=\lceil n/2\rceil}^{n}\Gamma\left(\left(\tfrac{pq}t\right)^{\frac{2(2n+1-2k)}{2n+3}}\right)}\,\frac{\kappa^n}{2^n}\oint\prod_{i=1}^n\frac{dz_i}{2\pi iz_i}\\
&\quad\times\prod_{i=1}^n\frac{\Gamma\left(z_i^{\pm2,0}\left(\tfrac{pq}t\right)^{\frac{2n+1-2i}{2n+3}}\right)}{\Gamma(z_i^{\pm2})}\,\Gamma\left(z_1^{\pm}\left(\tfrac{pq}t\right)^{\frac{1-2n}{2n+3}}\,t^{\frac12}\right)\,\Gamma\left(z_1^{\pm}\left(\tfrac{pq}t\right)^{\frac{2}{2n+3}}\,t^{\frac12}\right)\\
&\quad\times\prod_{i=2}^n\Gamma\left(z_{i-1}^{\pm}z_i^{\pm}\left(\tfrac{pq}t\right)^{\frac{4i-2n-1}{2(2n+3)}}\,t^{\frac12}\right)\,.\label{eq:gammaAeven}
\end{split}
\end{align}
On the other side of duality, the deformed $Sp(n)$ SQCD of~\cite{Maruyoshi:2016aim} gives 
\begin{align}
\begin{split}
&\CI^{(A_1, A_{2n})}_{Sp(n)}=\frac{\prod_{m=n+2}^{2n+1}\Gamma\left(\left(\tfrac{pq}t\right)^{\frac{2m}{2n+3}}\right)}{\prod_{i=1}^{n}\Gamma\left(\left(\tfrac{pq}t\right)^{\frac{2i}{2n+3}}\right)}\,\frac{\kappa^n}{2^nn!}\,\oint\prod_{i=1}^n\frac{dz_i}{2\pi iz_i}\prod_{i=1}^n\frac{\Gamma\left(z_i^{\pm2,0}\left(\tfrac{pq}t\right)^{\frac1{2n+3}}\right)}{\Gamma(z_i^{\pm2})}\\
&\qquad\times\prod_{1\le i<j\le n}\frac{\Gamma\left(z_i^{\pm}z_j^{\pm}\left(\tfrac{pq}t\right)^{\frac1{2n+3}}\right)}{\Gamma(z_i^{\pm}z_j^{\pm})}\times\prod_{i=1}^n\Gamma\left(z_i^{\pm}\left(\tfrac{pq}t\right)^{\frac{n+1}{2n+3}}t^{\frac12}\right)\,\Gamma\left(z_i^{\pm}\left(\tfrac{pq}t\right)^{-\frac{n}{2n+3}}t^{\frac12}\right)\,.\label{eq:gammaAevensqcd}
\end{split}
\end{align}
The duality implies that the RHS of \eqref{eq:gammaAeven} is equal to the RHS of  \eqref{eq:gammaAevensqcd}. 
Upon evaluating the integral for the quiver theory as a power series for $n=2, 3$, we obtain
\begin{align}
\begin{split}
    \CI_{(A_1,A_4)}&=1+\ft^{16/7}\xi^{16/7}+\ft^{20/7}\xi^{20/7}-\ft^{23/7}\xi^{2/7}\chi_{2}(y)-\ft^{27/7}\xi^{6/7}\chi_{2}(y)+\ft^{30/7}\xi^{-12/7}\\
&\quad+\ft^{32/7}\xi^{32/7}+\ft^{34/7}\xi^{-8/7}+\ft^{36/7}\xi^{36/7}+\ft^{37/7}\xi^{16/7}\chi_{2}(y)-\ft^{39/7}\xi^{18/7}\chi_{2}(y)\\
&\quad+\ft^{40/7}\xi^{40/7}+\ft^{41/7}\xi^{20/7}\chi_{2}(y)-\ft^{6}+\dots\,,\label{eq:fullA4}
\end{split}
\\
\begin{split}
\CI_{(A_1,A_6)}&=1+\ft^{20/9}\xi^{20/9}+\ft^{8/3}\xi^{8/3}+\ft^{28/9}\xi^{28/9}-\ft^{29/9}\xi^{2/9}\chi_{2}(y)-\ft^{11/3}\xi^{2/3}\chi_{2}(y)\\
&\quad-\ft^{37/9}\xi^{10/9}\chi_{2}(y)+\ft^{38/9}\xi^{-16/9}+\ft^{40/9}\xi^{40/9}+\ft^{14/3}\xi^{-4/3}+\ft^{44/9}\xi^{44/9}\\
&\quad+\ft^{46/9}\xi^{-8/9}+\ft^{47/9}\xi^{20/9}\chi_{2}(y)+2\ft^{16/3}\xi^{16/3}-\ft^{49/9}\xi^{22/9}\chi_{2}(y)\\
&\quad+\ft^{17/3}\xi^{8/3}\chi_{2}(y) +\ft^{52/9}\xi^{52/9}-2\ft^{53/9}\xi^{26/9}\chi_{2}(y)-\ft^{6}+\dots\,.\label{eq:fullA6}
\end{split}
\end{align}
This result agrees with the index obtained from the deformed $Sp(n)$ SQCD of~\cite{Maruyoshi:2016aim}. It would be desirable to find an analytic proof of the identity. 

In the Schur limit, the indices reduce to
\begin{align}
\CI_{(A_1,A_4)} &= 1+q^2+q^3+2q^4+2q^5+3q^6+3q^7+5q^8+\dots\,,\label{eq:schurA4}\\
\CI_{(A_1,A_6)} &= 1+q^2+q^3+2q^4+2q^5+4q^6+4q^7+6q^8+\dots\,,\label{eq:schurA6}
\end{align}
which agrees with the vacuum characters of the $(2,7)$ and $(2,9)$ Virasoro minimal models. 
This is consistent with the associated VOA for the $(A_1, A_{2n})$ theory being given by the $(2,2n+3)$ Virasoro minimal model.

\section{From \texorpdfstring{$SU(2)^n$}{SU(2)n} quiver to \texorpdfstring{$(A_1,D_{2n+2})$ and $(A_1,A_{2n+1})$}{(A1,D2n+2) and (A1,A2n+1)}}
\label{sec:D-even A-odd}

We now turn to theories with fundamentals at both ends of the quiver. The matter content is almost identical to the one we discussed in Section~\ref{sec:D-odd A-even}, except that we have two additional fundamentals $q_{n+1}$ of $SU(2)_n$ at the right end of the quiver. Two distinct choices of superpotential (and flip fields) lead the theory to flow to either $(A_1, D_{2n+2})$ or $(A_1, A_{2n+1})$ theory.

\subsection{\texorpdfstring{$(A_1,D_{2n+2})$ theory}{(A1,D2n+2)}}
\label{sec:D-even}

\begin{figure}[t]
\centering
\begin{tikzpicture}[gauge/.style={circle,draw,minimum size=7.5mm,inner sep=1pt},flavor/.style={rectangle,draw,minimum size=6.5mm,inner sep=3pt}]
\node[gauge] (g1) at (0,0) {$2$};
\node[gauge] (g2) at (1.9,0) {$2$};
\node (dd) at (3.8,0) {$\cdots$};
\node[gauge] (gm) at (5.7,0) {$2$};
\node[gauge] (gn) at (7.6,0) {$2$};
\node[flavor] (q0) at (-2.2,0) {$2$};
\draw (q0)-- node[above]{\footnotesize $q_1$} node[pos=0.5]{$\times$} node[below]{\footnotesize $M_1$} (g1);
\draw (g1)-- node[above]{\footnotesize $q_2$} node[pos=0.5]{$\times$} node[below]{\footnotesize $M_2$} (g2);
\draw (g2)-- node[above]{\footnotesize $q_3$} (dd);
\draw (dd)-- node[above]{\footnotesize $q_{n-1}$} (gm);
\draw (gm)-- node[above]{\footnotesize $q_n$} (gn);
\node[flavor] (qR) at (9.8,0) {$1$};
\draw ([yshift=2.5pt]gn.east)-- node[above]{\footnotesize $q_{n+1}$} ([yshift=2.5pt]qR.west);
\draw ([yshift=-2.5pt]qR.west)-- ([yshift=-2.5pt]gn.east);
\draw (g1) to[out=115,in=65,looseness=8] node[pos=0.5,above=3pt]{\footnotesize $\phi_1$} (g1);
\draw (g2) to[out=115,in=65,looseness=8] node[pos=0.5,above=3pt]{\footnotesize $\phi_2$} (g2);
\draw (gm) to[out=115,in=65,looseness=8] node[pos=0.5]{$\times$} node[pos=0.5,above=3pt]{\footnotesize $\phi_{n-1}$} node[pos=0.5,right=4pt]{\footnotesize $X_{n-1}$} (gm);
\draw (gn) to[out=115,in=65,looseness=8] node[pos=0.5]{$\times$} node[pos=0.5,above=3pt]{\footnotesize $\phi_n$} node[pos=0.5,right=4pt]{\footnotesize $X_n$} (gn);
\end{tikzpicture}
\caption{The quiver diagram for the $\CN=1$ theory flowing to $(A_1, D_{2n+2})$ theory. The cross on a line denotes a flip field: $M_i$ ($i\le\lceil n/2\rceil$) is a flip field for $q_i q_i$, and $X_k$ $(k\ge\lfloor n/2\rfloor+1)$ is a flip field for $\tr \phi_k^2$.}
\label{fig:D-even}
\end{figure}
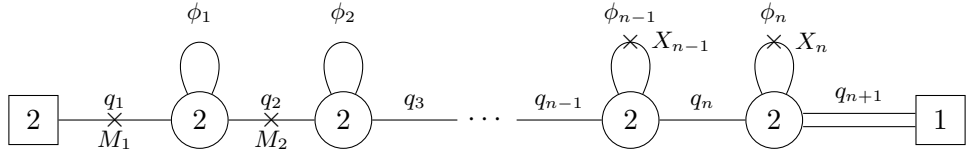

Consider the superpotential 
\begin{align}
W=\sum_{i=1}^{n-1}\phi_i q_{i+1}\phi_{i+1} q_{i+1} +q_{n+1}\phi_n q_{n+1}+\sum_{i=1}^{\lceil n/2\rceil}M_i q_i q_i +\sum_{k=\lfloor n/2\rfloor+1}^{n}X_k\tr\phi_k^2\,,\label{eq:WDA2} 
\end{align}
with the gauge-singlet flip fields $M_i$ and $X_k$. See Figure~\ref{fig:D-even} for a graphical presentation of our quiver gauge theory, and Table~\ref{tab:D-even} for the matter content and their charges.

\begin{table}[t]
\centering
\renewcommand{\arraystretch}{1.6}
\setlength{\tabcolsep}{5pt}
\begin{tabular}{llcc}
\hline
field & representation & $R$ & $f$\\
\hline
$\phi_i$, $1\le i\le n$ & $\mathbf 3$ of $SU(2)_i$ & $\dfrac{2(n+1-i)}{3(n+1)}$ & $\dfrac{2(n+1-i)}{n+1}$\\
$q_i$, $1\le i\le n$ & $(\mathbf 2_{i-1},\mathbf 2_{i})$ & $\dfrac{n+2i}{3(n+1)}$ & $-\dfrac{2n+3-2i}{n+1}$\\
$q_{n+1}$ & $\mathbf 2$ of $SU(2)_n$, $U(1)$ charges $\pm1$ & $\dfrac{3n+2}{3(n+1)}$ & $-\dfrac{1}{n+1}$\\
$M_i$, $1\le i\le\lceil\tfrac n2\rceil$ & $\mathbf 1$ & $\dfrac{2(2n+3-2i)}{3(n+1)}$ & $\dfrac{2(2n+3-2i)}{n+1}$\\
$X_k$, $\lfloor\tfrac n2\rfloor+1\le k\le n$ & $\mathbf 1$ & $2-\dfrac{4(n+1-k)}{3(n+1)}$ & $-\dfrac{4(n+1-k)}{n+1}$\\
\hline
\end{tabular}
\caption{Matter content of the $\CN=1$ theory that flows to $(A_1,D_{2n+2})$: gauge representation, infrared $R$-charge, and $U(1)_f$ charge of every field. $SU(2)_0$ is the flavor symmetry, and $U(1)$ the flavor symmetry under which the two fundamentals in $q_{n+1}$ have charges $\pm1$.}
\label{tab:D-even}
\end{table}

We proceed verbatim as in Section~\ref{sec:D-odd A-even}. We fix the superconformal $R$-charges from 1) the anomaly-free condition ($\tr R G_i G_i = 0$), 2) each term in the superpotential, and 3) the $a$-maximization procedure over one final free parameter. The resulting $R$-charges are given in Table~\ref{tab:D-even}, from which we compute the central charges to get
\begin{align}
(a,c)=\left(\frac n2+\frac1{12},\ \frac n2+\frac16\right)\,,
\end{align}
which is identical to that of the $(A_1,D_{2n+2})$ theory.

The Coulomb branch operators of $(A_1,D_{2n+2})$ theory correspond to flip fields $M_i$ with $i = 1, \cdots, \lceil n/2\rceil$, and also Casimir operators $\tr \phi_k^2$ with $k = 1, \cdots, \lfloor n/2\rfloor$. 
Their scaling dimensions are $\tfrac{2n+3-2i}{n+1}$ for $M_i$ and $\tfrac{2(n+1-k)}{n+1}$ for $\tr\phi_k^2$. These $n$ operators have dimensions $\{\tfrac m{n+1}\}_{m=n+2}^{2n+1}$, identical to the Coulomb-branch spectrum of the $(A_1,D_{2n+2})$ theory.

The theory \eqref{eq:WDA2} also has a flavor symmetry $SU(2)\times U(1)$, identical to that of $(A_1,D_{2n+2})$. The $SU(2)$ factor is given by $SU(2)_0$ in the quiver, rotating the two fundamentals $q_1$, and the $U(1)$ factor is the (baryonic) symmetry rotating the two fundamentals $q_{n+1}$ with charge $\pm 1$. Notice that the term $q_{n+1}^2\phi_n$ breaks the $SU(2)$ symmetry down to $U(1)$. 
The mesons $q_1\phi_1 q_1$ form an $SU(2)_0$ triplet with $\Delta=2$ and $f=-2$, which becomes the moment map for the $SU(2)$ flavor symmetry. The flavor central charge for $SU(2)$ is given by the trace anomaly
\begin{align}
    k_{SU(2)}= -6 \tr R F F = \frac{2(2n+1)}{n+1}\,,
\end{align}
which is precisely the value of the $(A_1,D_{2n+2})$ theory. 
Interestingly, we find the moment map for the $U(1)$ part is realized by the operator $q_{n/2+1}^2$ for even $n$ and by the flip field $X_{(n+1)/2}$ for odd $n$, which is the only other gauge-invariant operator that has $R=4/3, f=-2$. 

For $n=1$, the theory is given by simple $SU(2)$ gauge theory with four fundamentals $q_1,q_2$, one adjoint $\phi_1$, and two singlets (flip fields) $M_1$, $X_1$, with the superpotential $W = \phi_1q_2^2 + M_1 q_1^2+X_1\tr\phi_1^2$. This theory flows to $(A_1,D_4)=H_2 = D_2(SU(3))$ theory, which has $(a,c)=(\tfrac7{12},\tfrac23)$ and the flavor symmetry gets enhanced to $SU(3)$. This is identical to the one obtained via a nilpotent deformation of $\mathcal N=2$ $SU(2)$ SQCD~\cite{Agarwal:2016pjo}.

The superconformal index of our quiver gauge theory can be evaluated from an integral formula
\begin{align}
\begin{split}
\CI^{(A_1, D_{2n+2})}_{\text{quiver}}&=\frac{\prod_{i=1}^{\lceil n/2\rceil}\Gamma\left(\left(\tfrac{pq}t\right)^{\frac{2n+3-2i}{n+1}}\right)}{\prod_{k=\lfloor n/2\rfloor+1}^{n}\Gamma\left(\left(\tfrac{pq}t\right)^{\frac{2(n+1-k)}{n+1}}\right)}\,\frac{\kappa^n}{2^n}\oint\prod_{i=1}^n\frac{dz_i}{2\pi iz_i}\,\prod_{i=1}^n\frac{\Gamma\left(z_i^{\pm2,0}\left(\tfrac{pq}t\right)^{\frac{n+1-i}{n+1}}\right)}{\Gamma(z_i^{\pm2})}\\
&\quad\times\prod_{i=1}^n\Gamma\left(z_{i-1}^{\pm}z_i^{\pm}\left(\tfrac{pq}t\right)^{\frac{2i-n-2}{2(n+1)}}\,t^{\frac12}\right)\,\Gamma\left(z_n^{\pm}x^{\pm1}\left(\tfrac{pq}t\right)^{\frac{n}{2(n+1)}}\,t^{\frac12}\right)\,,\label{eq:gammaDeven}
\end{split}
\end{align}
where $z_0\equiv a$ and $x$ are the fugacities for the $SU(2)_0$ and $U(1)$ flavor symmetry respectively. 
The duality implies that the index is identical to the one computed using the deformed $SU(n+1)$ SQCD of~\cite{Agarwal:2016pjo}, which is given by an integral over the maximal torus of $SU(n+1)$
\begin{align}
\begin{split}
\CI^{(A_1, D_{2n+2})}_{SU(n+1)}&=\frac{\prod_{m=n+2}^{2n+1}\Gamma\left(\left(\tfrac{pq}t\right)^{\frac{m}{n+1}}\right)}{\prod_{i=2}^{n+1}\Gamma\left(\left(\tfrac{pq}t\right)^{\frac{i}{n+1}}\right)}\,\frac{\kappa^n}{(n+1)!}\,\Gamma\left(\left(\tfrac{pq}t\right)^{\frac1{n+1}}\right)^n\oint\prod_{i=1}^{n}\frac{dz_i}{2\pi iz_i}\\
&\quad\times\prod_{1\le i\ne j\le n+1}\frac{\Gamma\left(z_iz_j^{-1}\left(\tfrac{pq}t\right)^{\frac1{n+1}}\right)}{\Gamma(z_iz_j^{-1})}\prod_{i=1}^{n+1}\Gamma\left(\left(z_ia_1a_2^{2n+1}\right)^{\pm}\left(\tfrac{pq}t\right)^{\frac{n}{2(n+1)}}t^{\frac12}\right)\\
&\quad\times\prod_{i=1}^{n+1}\Gamma\left(\left(z_ia_1a_2^{-1}\right)^{\pm}\left(\tfrac{pq}t\right)^{-\frac{n}{2(n+1)}}t^{\frac12}\right)\,,\qquad\prod_{i=1}^{n+1}z_i=1\,,\label{eq:gammaDevensqcd}
\end{split}
\end{align}
where $a_1$ and $a_2$ are the fugacities of the $U(1)\times U(1)$ flavor symmetry of the SQCD description, which is enhanced to the $SU(2)\times U(1)$ flavor symmetry at the fixed point~\cite{Agarwal:2016pjo}. 
Evaluating the integral formula \eqref{eq:gammaDeven} for $n=2$ and $n=3$ as a power series gives
\begin{align}
\begin{split}
\CI_{(A_1,D_6)}&=1+\ft^{8/3}\xi^{8/3}+\ft^{10/3}\xi^{10/3}-\ft^{11/3}\xi^{2/3}\chi_{2}(y)+4\ft^{4}\xi^{-2}-\ft^{13/3}\xi^{4/3}\chi_{2}(y)\\
&\qquad+\ft^{14/3}\xi^{-4/3}+\ft^{16/3}\left(\xi^{16/3}+\xi^{-2/3}\right)+\ft^{17/3}\xi^{8/3}\chi_{2}(y) \\
&\qquad +\ft^{6}\left(\xi^{6}-5+4\xi^{-3}\right)+\dots\,,\label{eq:fullD6}
\end{split}
\\
\begin{split}
\CI_{(A_1,D_8)}&=1+\ft^{5/2}\xi^{5/2}+\ft^{3}\xi^{3}+\ft^{7/2}\left(\xi^{7/2}-\xi^{1/2}\chi_{2}(y)\right)+\ft^{4}\left(-\xi\chi_{2}(y)+4\xi^{-2}\right)\\
&\qquad+\ft^{9/2}\left(-\xi^{3/2}\chi_{2}(y)+\xi^{-3/2}\right)+\ft^{5}\left(\xi^{5}+\xi^{-1}\right) \\
&\qquad +\ft^{11/2}\left(\xi^{11/2}+\xi^{5/2}\chi_{2}(y)+\xi^{-1/2}\right)
+\ft^{6}\left(2\xi^{6}-5\right)+\dots\,, \label{eq:fullD8}
\end{split}
\end{align}
where we turned off the flavor fugacities for simplicity. 
They indeed agree with the ones obtained from the deformed $SU(n+1)$ SQCD of~\cite{Agarwal:2016pjo}.\footnote{The $(A_1,D_6)$ index presented in~\cite{Agarwal:2016pjo} has a typo at order $\ft^{23/3}$. The notations map as $v_{\text{there}}=\xi_{\text{here}}^2$, and $z_{\text{there}}$ being the fugacity for the $SU(2)$ flavor symmetry. The term $-v^{2/3}\chi_2(y)\left(\chi_3(y)-1\right)$ should be read as $-v^{-2/3}\chi_2(y)\left(\chi_3(z)-1\right)$.}

We can take the Schur limit of the index to obtain
\begin{align}
I_{(A_1, D_6)} &= 1+4q+4q^{3/2}+14q^2+16q^{5/2}+46q^3+56q^{7/2}+129q^4+168q^{9/2}%+336q^5
+\dots\,,\label{eq:schurD6}\\
I_{(A_1, D_8)} &= 1+4q+18q^2+56q^3+167q^4+436q^5+1086q^6+2520q^7+5631q^8+\dots\,,\label{eq:schurD8}
\end{align}
which are the vacuum characters of $\mathcal W_{-8/3}(\mathfrak{sl}_4,[2,1^2])$ and $\mathcal W_{-15/4}(\mathfrak{sl}_5,[3,1^2])$. This is consistent with the associated VOA for $(A_1,D_{2n+2})$ being given by $\CW_{-\frac{n(n+2)}{n+1}}(\mathfrak{sl}_{n+2},[n,1^2])$~\cite{Creutzig:2017qyf,Song:2017oew}.

\subsection{\texorpdfstring{$(A_1,A_{2n+1})$ theory}{(A1,A2n+1)}}
\label{sec:A-odd}
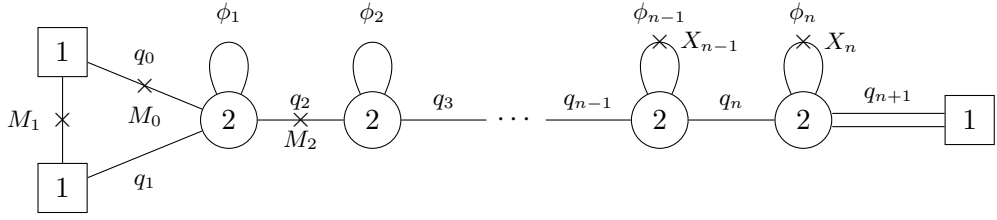
\begin{figure}[t]
\centering
\begin{tikzpicture}[gauge/.style={circle,draw,minimum size=7.5mm,inner sep=1pt},flavor/.style={rectangle,draw,minimum size=6.5mm,inner sep=3pt}]
\node[gauge] (g1) at (0,0) {$2$};
\node[gauge] (g2) at (1.9,0) {$2$};
\node (dd) at (3.8,0) {$\cdots$};
\node[gauge] (gm) at (5.7,0) {$2$};
\node[gauge] (gn) at (7.6,0) {$2$};
\node[flavor] (qa) at (-2.2,0.9) {$1$};
\node[flavor] (qb) at (-2.2,-0.9) {$1$};
\draw (qa)-- node[pos=0.5]{$\times$} node[pos=0.5,above=4pt]{\footnotesize $q_0$} node[pos=0.5,below=4pt]{\footnotesize $M_0$} (g1);
\draw (qb)-- node[pos=0.5,below=4pt]{\footnotesize $q_1$} (g1);
\draw (qa)-- node[pos=0.5]{$\times$} node[pos=0.5,left=4pt]{\footnotesize $M_1$} (qb);
\draw (g1)-- node[above]{\footnotesize $q_2$} node[pos=0.5]{$\times$} node[below]{\footnotesize $M_2$} (g2);
\draw (g2)-- node[above]{\footnotesize $q_3$} (dd);
\draw (dd)-- node[above]{\footnotesize $q_{n-1}$} (gm);
\draw (gm)-- node[above]{\footnotesize $q_n$} (gn);
\node[flavor] (qR) at (9.8,0) {$1$};
\draw ([yshift=2.5pt]gn.east)-- node[above]{\footnotesize $q_{n+1}$} ([yshift=2.5pt]qR.west);
\draw ([yshift=-2.5pt]qR.west)-- ([yshift=-2.5pt]gn.east);
\draw (g1) to[out=115,in=65,looseness=8] node[pos=0.5,above=3pt]{\footnotesize $\phi_1$} (g1);
\draw (g2) to[out=115,in=65,looseness=8] node[pos=0.5,above=3pt]{\footnotesize $\phi_2$} (g2);
\draw (gm) to[out=115,in=65,looseness=8] node[pos=0.5]{$\times$} node[pos=0.5,above=3pt]{\footnotesize $\phi_{n-1}$} node[pos=0.5,right=4pt]{\footnotesize $X_{n-1}$} (gm);
\draw (gn) to[out=115,in=65,looseness=8] node[pos=0.5]{$\times$} node[pos=0.5,above=3pt]{\footnotesize $\phi_n$} node[pos=0.5,right=4pt]{\footnotesize $X_n$} (gn);
\end{tikzpicture}
\caption{The quiver diagram for the $\CN=1$ theory flowing to $(A_1, A_{2n+1})$ theory. The cross on a line denotes a flip field: $M_i$ $(2\le i\le\lfloor n/2\rfloor)$ is a flip field for $q_i q_i$, and $X_k$ $(k\ge\lceil n/2\rceil)$ is a flip field for $\tr \phi_k^2$.}
\label{fig:A-odd}
\end{figure}

Now, let us consider the same quiver, but with an interaction that breaks $SU(2)_0$. We write these two fundamentals of $SU(2)_1$ as $q_0$ and $q_1$. See Table \ref{tab:A-odd} for the matter content and their charges. 
We write the superpotential as
\begin{align}
\begin{split}
    W &=\sum_{i=1}^{n-1}\phi_i q_{i+1} \phi_{i+1} q_{i+1} + q_1 \phi_1 q_1 + q_{n+1}\phi_nq_{n+1} \\
    &\quad +M_0 q_0\phi_1q_0+M_1 q_0q_1  +\sum_{i=2}^{\lfloor n/2\rfloor}M_iq_iq_i + \sum_{k=\lceil n/2\rceil}^{n}X_k\tr\phi_k^2\,.\label{eq:WA1A2}
\end{split}
\end{align}
As before, we have cubic interactions for the fundamentals at both ends of the quiver, quartic interactions for the bifundamentals, and flip fields for the (generalized) mesonic operators and Casimir operators. See Figure~\ref{fig:A-odd} for a graphical presentation of our quiver gauge theory. 

\begin{table}[t]
\centering
\renewcommand{\arraystretch}{1.6}
\setlength{\tabcolsep}{5pt}
\begin{tabular}{llcc}
\hline
field & representation & $R$ & $f$\\
\hline
$\phi_i$, $1\le i\le n$ & $\mathbf 3$ of $SU(2)_i$ & $\dfrac{2(n+1-i)}{3(n+2)}$ & $\dfrac{2(n+1-i)}{n+2}$\\
$q_0$ & $\mathbf 2$ of $SU(2)_1$ & $\dfrac{4}{3(n+2)}$ & $-\dfrac{3n+2}{n+2}$\\
$q_1$ & $\mathbf 2$ of $SU(2)_1$ & $\dfrac{2n+6}{3(n+2)}$ & $-\dfrac{n}{n+2}$\\
$q_i$, $2\le i\le n$ & $(\mathbf 2_{i-1},\mathbf 2_{i})$ & $\dfrac{n+2i+3}{3(n+2)}$ & $-\dfrac{2n+3-2i}{n+2}$\\
$q_{n+1}$ & $\mathbf 2$ of $SU(2)_n$, $U(1)$ charges $\pm1$ & $\dfrac{3n+5}{3(n+2)}$ & $-\dfrac{1}{n+2}$\\
$M_0$ & $\mathbf 1$ & $\dfrac{4(n+1)}{3(n+2)}$ & $\dfrac{4(n+1)}{n+2}$\\
$M_1$ & $\mathbf 1$ & $\dfrac{2(2n+1)}{3(n+2)}$ & $\dfrac{2(2n+1)}{n+2}$\\
$M_i$, $2\le i\le\lfloor\tfrac{n}{2}\rfloor$ & $\mathbf 1$ & $\dfrac{2(2n+3-2i)}{3(n+2)}$ & $\dfrac{2(2n+3-2i)}{n+2}$\\
$X_k$, $\lceil\tfrac{n}{2}\rceil\le k\le n$ & $\mathbf 1$ & $2-\dfrac{4(n+1-k)}{3(n+2)}$ & $-\dfrac{4(n+1-k)}{n+2}$\\
\hline
\end{tabular}
\caption{Matter content of the $\CN=1$ theory that flows to $(A_1,A_{2n+1})$: gauge representation, infrared $R$-charge, and $U(1)_f$ charge of every field. $U(1)$ is the flavor symmetry under which the two fundamentals in $q_{n+1}$ have charges $\pm1$.}
\label{tab:A-odd}
\end{table}

We perform the $a$-maximization procedure as before, and the resulting $R$-charges are written in Table~\ref{tab:A-odd}. The resulting central charges are given as
\begin{align}
(a,c)=\left(\frac{12n^2+19n+2}{24(n+2)},\ \frac{3n^2+5n+1}{6(n+2)}\right)\,,
\end{align}
which precisely agree with the values of $(A_1,A_{2n+1})$ theory.

The Coulomb branch of the $(A_1, A_{2n+1})$ theory is realized by the flip fields $M_{i = 0, 1, \cdots, \lfloor n/2\rfloor}$ for the (generalized) mesons, and the (unflipped) Casimir operators $\tr\phi_{k=1, \cdots, \lceil n/2\rceil-1}^2$. 
Their scaling dimensions are $\Delta(M_0) = \tfrac{2(n+1)}{n+2}$, $\Delta(M_{i\ge 1}) = \tfrac{2n+3-2i}{n+2}$ and $\Delta(\tr\phi_k^2) = \tfrac{2(n+1-k)}{n+2}$. 
These $n$ operators have dimensions $\{\tfrac m{n+2}\}_{m=n+3}^{2n+2}$, agreeing with the Coulomb-branch spectrum of the $(A_1,A_{2n+1})$ theory.

The theory \eqref{eq:WA1A2} has a $U(1)$ flavor symmetry (enhanced to $SU(2)$ for $n=1$), which is the same as that of $(A_1, A_{2n+1})$ theory, under which the two fundamentals $q_{n+1}$ have charges $\pm1$. Its moment map is realized by the operator $q_{(n+1)/2}^2$ for odd $n$ and by the flip field $X_{n/2}$ for even $n$. 

We can understand this theory from the RG flow $(A_1,D_{2n+2})\to(A_1,A_{2n+1})$ triggered by nilpotent Higgsing of the flip field for the moment map~\cite{Maruyoshi:2016aim}. The operator $q_1^2\phi_1$ is a component of the moment-map in $(A_1,D_{2n+2})$ theory, which has $\D=2$ and $f=-2$. Adding the (triplet) flip deformation $M q_0\phi_1 q_0$ to \eqref{eq:WDA2} and then Higgsing $\vev{M} = \sigma^+$ triggers flow to \eqref{eq:WA1A2}. 

When $n=1$, the theory is given by a simple $SU(2)$ gauge theory with four fundamentals, one adjoint $\phi_1$, and the singlets $M_0$, $M_1$, and $X_1$. Now $M_1$ saturates the unitarity bound (and becomes free), so we remove it. The remaining theory is identical to the one studied in \cite{Agarwal:2016pjo}, which is obtained via nilpotent Higgsing of $SU(2)$ SQCD. (It is referred to as $(A_1, D_3)$ in \cite{Agarwal:2016pjo}.) This theory indeed flows to $(A_1,A_3)=(A_1, D_3) = H_1 = D_3(SU(2))$ theory, which has central charges $(a,c)=(\tfrac{11}{24},\tfrac12)$ and a Coulomb branch operator of dimension $\Delta = 4/3$. 

The superconformal index of the quiver theory takes the following integral form,
\begin{align}
\begin{split}
\CI^{(A_1, A_{2n+1})}_{\text{quiver}}&=\frac{\Gamma\left(\left(\tfrac{pq}t\right)^{\frac{2(n+1)}{n+2}}\right)\,\Gamma\left(\left(\tfrac{pq}t\right)^{\frac{2n+1}{n+2}}\right)\prod_{i=2}^{\lfloor n/2\rfloor}\Gamma\left(\left(\tfrac{pq}t\right)^{\frac{2n+3-2i}{n+2}}\right)}{\prod_{k=\lceil n/2\rceil}^{n}\Gamma\left(\left(\tfrac{pq}t\right)^{\frac{2(n+1-k)}{n+2}}\right)}\,\frac{\kappa^n}{2^n}\oint\prod_{i=1}^n\frac{dz_i}{2\pi iz_i}\\
&\quad\times\prod_{i=1}^n\frac{\Gamma\left(z_i^{\pm2,0}\left(\tfrac{pq}t\right)^{\frac{n+1-i}{n+2}}\right)}{\Gamma(z_i^{\pm2})}\,\Gamma\left(z_1^{\pm}\left(\tfrac{pq}t\right)^{-\frac{n}{n+2}}\,t^{\frac12}\right)\,\Gamma\left(z_1^{\pm}\left(\tfrac{pq}t\right)^{\frac{1}{n+2}}\,t^{\frac12}\right)\\
&\quad\times\prod_{i=2}^n\Gamma\left(z_{i-1}^{\pm}z_i^{\pm}\left(\tfrac{pq}t\right)^{\frac{2i-n-1}{2(n+2)}}\,t^{\frac12}\right)\,\Gamma\left(z_n^{\pm}x^{\pm1}\left(\tfrac{pq}t\right)^{\frac{n+1}{2(n+2)}}\,t^{\frac12}\right)\,,\label{eq:gammaAodd}
\end{split}
\end{align}
where $x$ is the fugacity for the $U(1)$ flavor symmetry. 
In the dual $SU(n+1)$ Lagrangian description~\cite{Maruyoshi:2016aim}, the same index is given as an integral over the maximal torus of $SU(n+1)$
\begin{align}
\begin{split}
\CI^{(A_1, A_{2n+1})}_{SU(n+1)}&=\frac{\prod_{m=n+3}^{2n+2}\Gamma\left(\left(\tfrac{pq}t\right)^{\frac{m}{n+2}}\right)}{\prod_{i=2}^{n+1}\Gamma\left(\left(\tfrac{pq}t\right)^{\frac{i}{n+2}}\right)}\,\frac{\kappa^n}{(n+1)!}\,\Gamma\left(\left(\tfrac{pq}t\right)^{\frac1{n+2}}\right)^n\oint\prod_{i=1}^{n}\frac{dz_i}{2\pi iz_i}\\
&\quad\times\prod_{1\le i\ne j\le n+1}\frac{\Gamma\left(z_iz_j^{-1}\left(\tfrac{pq}t\right)^{\frac1{n+2}}\right)}{\Gamma(z_iz_j^{-1})}\prod_{i=1}^{n+1}\Gamma\left(\left(z_ix\right)^{\pm}\left(\tfrac{pq}t\right)^{-\frac{n}{2(n+2)}}t^{\frac12}\right)\,,\label{eq:gammaAoddsqcd}
\end{split}
\end{align}
with the constraint $\prod_{i=1}^{n+1}z_i=1$. 
The duality implies that the RHS of \eqref{eq:gammaAodd} equals the RHS of \eqref{eq:gammaAoddsqcd}. 
Upon evaluating the integral \eqref{eq:gammaAodd} as a power series for $n=2$ and $n=3$, we obtain
\begin{align}
\begin{split}
\CI_{(A_1,A_5)}&=1+\ft^{5/2}\xi^{5/2}+\ft^{3}\xi^{3}-\ft^{7/2}\xi^{1/2}\chi_{2}(y)+\ft^{4}\left(-\xi\chi_{2}(y)+\xi^{-2}\right)+\ft^{9/2}\xi^{-3/2}\\
&\quad+\ft^{5}\left(\xi^{5}+\xi^{-1}\right)+\ft^{11/2}\left(\xi^{11/2}+\xi^{5/2}\chi_{2}(y)\right)+\ft^{6}\left(\xi^{6}-2+2\xi^{-3}\right)+\dots\,,\label{eq:fullA5}
\end{split}
\\
\begin{split}
\CI_{(A_1,A_7)}&=1+\ft^{12/5}\xi^{12/5}+\ft^{14/5}\xi^{14/5}+\ft^{16/5}\xi^{16/5}-\ft^{17/5}\xi^{2/5}\chi_{2}(y)-\ft^{19/5}\xi^{4/5}\chi_{2}(y)\\
&\quad+\ft^{4}\xi^{-2}-\ft^{21/5}\xi^{6/5}\chi_{2}(y)+\ft^{22/5}\xi^{-8/5}+\ft^{24/5}\left(\xi^{24/5}+\xi^{-6/5}\right)\\
&\quad +\ft^{26/5}\left(\xi^{26/5}+\xi^{-4/5}\right)+\ft^{27/5}\xi^{12/5}\chi_{2}(y)+2\ft^{28/5}\xi^{28/5}+\ft^{6}\left(\xi^{6}-2\right)+\dots\,,\label{eq:fullA7} 
\end{split}
\end{align}
where we turned off the flavor fugacity for simplicity. 
These indices indeed agree with the ones obtained from the deformed $SU(n+1)$ SQCD of \cite{Maruyoshi:2016aim}.
Taking the Schur limit of the index, we obtain
\begin{align}
\CI_{(A_1,A_5)}&=1+q+2q^{3/2}+3q^2+4q^{5/2}+7q^3+8q^{7/2}+14q^4+18q^{9/2}%+26q^5
+\dots\,,\label{eq:schurA5}\\
\CI_{(A_1,A_7)}&=1+q+5q^2+9q^3+20q^4+36q^5+71q^6+119q^7+213q^8+\dots\,,\label{eq:schurA7}
\end{align}
which agrees with the Schur index computed in \cite{Buican:2015ina}. They should be identical to the vacuum characters of $\mathcal W_{-9/4}(\mathfrak{sl}_3,[2,1])$ and $\mathcal W_{-16/5}(\mathfrak{sl}_4,[3,1])$. More generally, the associated VOA for $(A_1, A_{2n+1})$ is conjectured to be given by $\CW_{-\frac{(n+1)^2}{n+2}}(\mathfrak{sl}_{n+1},[n,1])$~\cite{Creutzig:2017qyf, Xie:2016evu, Song:2017oew}, which is a Drinfeld-Sokolov reduction of affine Kac-Moody algebra via subregular orbit.

\section{Discussion}
In this paper, we proposed a family of new $\CN=1$ Lagrangian $SU(2)^n$ quiver gauge theories that are infrared dual to the ones given by $Sp(n)/SU(n+1)$ gauge theories \cite{Maruyoshi:2016aim, Agarwal:2016pjo} that flow to the $\CN=2$ Argyres-Douglas theories. We have performed various computations, including the full superconformal index, to test the duality. 

One notable feature of the new dual is that Coulomb branch operators come from two sets of gauge-invariant operators: one set from the gauge-singlet fields $ M_i$ for the mesons $q_i q_i$, and the other set from the Casimir operators $\tr \phi_i^2$. This feature is different from the $SU(n+1)/Sp(n)$ description, where all the Coulomb branch operators are given by the flip fields for the (generalized) mesons. On the other hand, this is reminiscent of the $\CN=1$ quiver theory descriptions for $(A_{k-1}, A_{mk-1})$ and others \cite{Agarwal:2017roi, Benvenuti:2017bpg}. 

Another characteristic feature that differs from the previous $\CN=1$ Lagrangians is that it does not have an obvious $\CN=2$ origin. In \cite{Maruyoshi:2016aim, Agarwal:2016pjo}, such gauge theories were obtained via $\CN=1$ -- preserving deformations of $\CN=2$ SQCD. The new description does not obviously arise from $\CN=2$, as can be seen from the fact that we only have a \emph{single} bifundamental between the gauge nodes, instead of a pair of bifundamentals which more naturally arise in string/M-theoretic setups. On the other hand, it might still be possible to obtain our gauge theory from compactification of a 6d theory \cite{Gaiotto:2009hg, Gaiotto:2009we}, as was the case of the previous $SU/Sp$ description \cite{Xie:2012hs, Maruyoshi:2016aim, Giacomelli:2017ckh, Carta:2018qke, Carta:2019hbi}. It is actually possible to obtain the left-hand side of our quiver from the $\CN=1$ class $\mathcal{S}$ setup \cite{Agarwal:2014rua, Agarwal:2015vla, Hwang:2021xyw}. It would be interesting to look for a geometric realization of the quiver description, which may give hints towards deriving the duality.

One possible hint of this duality comes from five dimensions. The 5d $\mathcal N=1$ $SU(n+1)$ gauge theories with $2n+2$ flavors and the $SU(2)^{n}$ linear quiver gauge theory with 2 flavors at its two ends are two dual descriptions for the same 5d SCFT. They are related by the $S$-duality of the $(p,q)$ 5-brane web that realizes them --- a rotation of the web by $90^\circ$, which exchanges the number of parallel D5-branes with the number of parallel NS5-branes~\cite{Aharony:1997ju, Aharony:1997bh}, or the exchange of the base and the fiber of the Calabi--Yau geometry that engineers them~\cite{Katz:1997eq, Bao:2011rc}. There too, the total rank is preserved, and the rank of the simple gauge group becomes the length of the $SU(2)$ quiver. The 5d duality does not reduce to the one found here: compactified to four dimensions with $\mathcal N=2$ supersymmetry it becomes the S-duality of $SU(n+1)$ with $2n+2$ flavors, in which all but one of the $SU(2)$ factors are replaced by the isolated theory $R_{0,n+1}$~\cite{Gaiotto:2009we, Chacaltana:2010ks, Bergman:2014kza}, and no Lagrangian $SU(2)^{n}$ quiver survives. It would be interesting to see whether we can understand our duality from a five-dimensional picture.

A more promising route is to consider the three-dimensional reduction, where the Argyres-Douglas theories themselves have Lagrangian descriptions. Their 3d mirrors are abelian $\mathcal N=4$ theories \cite{Nanopoulos:2010bv, Xie:2012hs, Buican:2015hsa, Benvenuti:2017kud, Benvenuti:2018bav, Dedushenko:2019mnd}: $(A_1, D_{2n+2})$ reduces to a linear quiver of $n$ $U(1)$ nodes with two flavors at one end and one at the other, and $(A_1,A_{2n+1})$ to a linear quiver of $n$ $U(1)$ nodes with one flavor at each end; $(A_1,D_{2n+1})$ reduces to $U(1)$ with two flavors together with $n-1$ free twisted hypermultiplets, and $(A_1,A_{2n})$ to $n$ free twisted hypermultiplets. For the SQCD Lagrangians of $(A_1,A_{2n+1})$ and $(A_1,D_{2n+2})$ this reduction has been carried out on the 3d mirror of the reduced SQCD: there the nilpotent deformation becomes a monopole superpotential that confines the nodes of the mirror one by one, down to the mirror of the abelian theory above \cite{Benvenuti:2017kud, Benvenuti:2017bpg}. We should be able to analyze our quiver gauge theories under dimensional reduction in a similar way to arrive more directly at the expected 3d mirror theory. 
Relatedly, we expect that our dual quiver theory provides an alternative way to perform the $U(1)_r$-twisted dimensional reduction \cite{Dedushenko:2023cvd, Gaiotto:2024ioj, ArabiArdehali:2024ysy} to obtain the theory of \cite{Gang:2018huc, Gang:2023rei, ArabiArdehali:2024vli, Kim:2025klh} whose boundary VOA upon topological twisting becomes the same as the associated VOA of the 4d theory.

\begin{acknowledgments}
We thank Sungjoon Kim and Shlomo Razamat for useful discussions.
This work is supported by the National Research Foundation of Korea (NRF) grants RS-2024-00405629 and RS-2026-25482546, the KAIST-KIAS collaboration program, and a KIAS Individual Grant PG111301 at Korea Institute for Advanced Study (MC). 
\end{acknowledgments}

\appendix

\section{Methods and Conventions} \label{sec:methods}
In this appendix, we describe the methods we used to discover our dual $\CN=1$ quiver gauge theories. Following the procedure of the authors' previous work \cite{Maruyoshi:2018nod, Cho:2024civ}, we performed an extensive scan of superconformal field theories that can be obtained from quiver gauge theories of low rank. The scanning procedure proceeds as follows: Start from a `seed' SCFT that is realized as an infrared fixed point of a gauge theory (in this case $SU(2) \times SU(2)$ with a single bifundamental, two adjoints, and a number of fundamental chiral multiplets for each factor) without superpotential. Then enumerate all the relevant gauge-invariant operators, and perform all possible deformations triggered by them. In addition, we also consider flip deformations for the `super-relevant' operators that have $R<4/3$ \cite{Barnes:2004jj}. This triggers an RG flow to another fixed point, which we can analyze via $a$-maximization and the superconformal index. We test unitarity and flip any possible unitarity-violating operators \cite{Kutasov:2003iy, Benvenuti:2017lle, Maruyoshi:2018nod} along the flow to isolate the interacting SCFT. Once we obtain a new fixed point, we repeat the procedure. This procedure eventually terminates once it lands on a theory where supersymmetric deformation no longer creates a new CFT. It yields a large landscape of SCFTs. Among them, we search for fixed points with rational central charges, which is a necessary condition for the theory to be $\CN=2$ supersymmetric \cite{Maruyoshi:2016aim, Rastelli:2023sfk} and compare that with the known AD theories. 

In the procedure, determining the superconformal $R$-symmetry is the most crucial part. It is determined by $a$-maximization \cite{Intriligator:2003jj}, combined with the anomaly-free condition for the $R$-symmetry. 
We start with a trial $R$-charge that is anomaly-free and compatible with the superpotential. For a gauge theory with gauge group $\prod_g G_g$ and chiral multiplets $\ch$ in the representations of dimension $|\ch|$, we have
\begin{align}\label{eq:anomalies}
    \tr R^3 = \sum_g |G_g| + \sum_\ch |\ch| (R_\ch-1)^3 \,, \quad \tr R = \sum_g |G_g| + \sum_\ch |\ch| (R_\ch-1)\,. 
\end{align}
The trial $R$-symmetry is anomaly-free when $\tr RGG$ vanishes at every gauge node,
\begin{align}
    \tr RG_gG_g = h_g^\vee + \sum_\ch T_g(\ch)(R_\ch-1)=0\,,
\end{align}
with $T_g(\ch)$ being the Dynkin index of $\ch$ under $G_g$. We choose the standard normalization of $\tr T^A T^B |_{\text{fund}} = 1/2$ so that a bifundamental chiral multiplet $(\bm 2_i , \bm 2_{i+1})$ contributes $T_i=2\cdot\half=1$.
Then, the superconformal $R$-symmetry should locally maximize the trial central charge $a(R)$, which is given in terms of the trace anomalies as \cite{Anselmi:1997am}
\begin{align}\label{eq:ac}
    a=\frac{3}{32}\left(3\tr R^3 - \tr R\right)\,, \qquad c= \frac{1}{32}\left(9 \tr R^3 - 5\tr R\right)\,.
\end{align}
The superconformal $R$-charge is fixed by imposing
\begin{align}
    \frac{\partial a(R_{\text{tr}})}{\partial R_{\text{tr}}} = 0, \quad \frac{\partial^2 a}{\partial R_{\text{tr}}^2} < 0 \ . 
\end{align}

We test our dualities by computing the superconformal index. The superconformal index \cite{Romelsberger:2005eg,Kinney:2005ej} can be defined in terms of a trace formula
\begin{align}
\CI(\ft,y;\xi)=\mathrm{Tr}\,(-1)^{F}\,\ft^{3(R+2j_1)}\,y^{2j_2}\,\xi^{f}\,\prod_i a_i^{F_i}\,,\qquad p=\ft^3y\,,\quad q=\ft^3/y\,,
\end{align}
where the trace is taken over certain supersymmetric states. 
Here $R$ is the superconformal $U(1)_R$ charge, $(j_1,j_2)$ are the spins of the $SU(2)\times SU(2)$ Lorentz group, and $f$ is the charge under a $U(1)$ global symmetry $U(1)_f$, which is a part of $\CN=2$ R-symmetry as we describe below, and $F_i$ are the Cartan generators of the flavor symmetry. 
Once the superconformal $R$ charge for each chiral multiplet is known, the index can be computed from the ultraviolet field content. A chiral multiplet with gauge character $\chi$, flavor weight $w$ and $U(1)_f$ charge $f$, and the vector multiplet contribute to the single-letter indices as
\begin{align}
i_\chi=\frac{\ft^{3R}\chi(z)\,a^w\,\xi^{f}-\ft^{3(2-R)}\chi(z^{-1})\,a^{-w}\,\xi^{-f}}{(1-\ft^3y)(1-\ft^3/y)}\,,\qquad i_V=\frac{2\ft^6-\ft^3(y+1/y)}{(1-\ft^3y)(1-\ft^3/y)}\,\chi_{\rm adj}(z)\,. 
\end{align}
Now, take the plethystic exponential and integrate over the gauge holonomy to obtain the superconformal index:
\begin{align}
\mathcal I=\oint d\mu(z)\,\mathrm{PE}\left[i_V+\sum_\chi i_\chi\right]\,,\qquad \mathrm{PE}[g](\ft,y,\xi,a,z)=\exp\sum_{k\ge1}\frac1k\,g(\ft^k,y^k,\xi^k,a^k,z^k)\,,
\label{eq:indexdef}
\end{align}
where $d\m$ is the Haar measure of the gauge group.
The index \eqref{eq:indexdef} can be rewritten in terms of a contour integral of elliptic gamma functions as \cite{Dolan:2008qi},
\begin{align}
    \G(z) \equiv \prod_{m,n\ge 0} \frac{1-z^{-1}p^{m+1}q^{n+1}}{1-zp^mq^n}\,,\quad \k\equiv(p;p)(q;q)\,,\quad (z;q)=\prod_{m\ge0}(1-zq^m)\,,
\end{align}
where we use the short-hand notation $f(z^\pm)\equiv f(z^+)f(z^-)$. The plethystic exponential of $i_\chi$ gives $\prod_{\rho}\Gamma\left((pq)^{R/2}\xi^{f}z^{\rho}\, a^{\,w}\right)$, with $\rho$ running over the weights of the gauge group representation. The contribution of the vector multiplets and the Haar measure is given as $\k^{\mathrm{rank}G}|W|^{-1}\prod_{\a} \G(z^\a)^{-1}$, where $W$ is the Weyl group, and $\a$ runs over the roots of the gauge group. 
The index \eqref{eq:indexdef} is then
\begin{align}
\CI=\frac{\kappa^{\mathrm{rank}\,G}}{|W|}\oint\prod_{i=1}^{\mathrm{rank}\,G}\frac{dz_i}{2\pi iz_i}\,\prod_{\alpha}\Gamma\left(z^{\alpha}\right)^{-1}\prod_{\chi}\prod_{\rho}\Gamma\left((pq)^{R_\chi/2}\,\xi^{f_\chi}\,z^{\rho}\,a^{\,w_\chi}\right)\,.\label{eq:indexgamma}
\end{align}
As we discussed in the main text, our duality predicts new identities among the integrals of elliptic Gamma functions. 

If the infrared theory has $\mathcal N=2$ supersymmetry, the $\mathcal N=1$ $R$-symmetry and $U(1)_f$ get enhanced to $\CN=2$ $SU(2)_R\times U(1)_r$ R-symmetry. The relations among Cartans are given as $R=\tfrac13r+\tfrac43I_3$ and $f=r-2I_3$, where $I_3$ is the $SU(2)_R$ Cartan and $r$ the $U(1)_r$ charge. This implies $f=+2$ for the would-be $\mathcal N=2$ vector multiplet adjoint, $f=-1$ for the hypermultiplets, $f=3R$ for the Coulomb-branch operators, and $f=-2$ for the moment maps. 
Each of the four quiver theories we consider has an anomaly-free $U(1)_f$, preserving its superpotential. Its charges are listed in the tables of Sections~\ref{sec:D-odd}--\ref{sec:A-odd}, where $f=3R$ for all the Coulomb-branch operators and $f=-2$ for all the moment map operators we identify. 
With $R$ and $f$ embedded as such, the trace formula can be written as a $\mathcal N=2$ superconformal index
\begin{align}
\mathcal I=\mathrm{Tr}\,(-1)^{F}\,p^{\,j_1+j_2+\frac r2}\;q^{\,j_1-j_2+\frac r2}\;t^{\,I_3-\frac r2}\,,\qquad t\equiv\mathfrak t^4/\xi^2\,.
\end{align}
The Schur limit is the specialization $t\to q$~\cite{Gadde:2011uv}, in which the $p$-dependence is automatically dropped.

%%%%%%%%%%%%%%%%%%%%%%%%%%%%%%%%%%%%%%%%%%
\bibliographystyle{jhep}
\bibliography{refs}

@article{Wang:2015mra,
      author         = "Wang, Yifan and Xie, Dan",
      title          = "{Classification of Argyres-Douglas theories from M5
                        branes}",
      journal        = "Phys. Rev.",
      volume         = "D94",
      year           = "2016",
      number         = "6",
      pages          = "065012",
      doi            = "10.1103/PhysRevD.94.065012",
      eprint         = "1509.00847",
      archivePrefix  = "arXiv",
      primaryClass   = "hep-th",
      reportNumber   = "MIT-CTP-4711",
      SLACcitation   = "%%CITATION = ARXIV:1509.00847;%%"
}

@article{Carta:2018qke,
      author         = "Carta, Federico and Giacomelli, Simone and Savelli,
                        Raffaele",
      title          = "{SUSY enhancement from T-branes}",
      journal        = "JHEP",
      volume         = "12",
      year           = "2018",
      pages          = "127",
      doi            = "10.1007/JHEP12(2018)127",
      eprint         = "1809.04906",
      archivePrefix  = "arXiv",
      primaryClass   = "hep-th",
      reportNumber   = "ROM2F-2018-05, IFT-UAM-CSIC-18-93",
      SLACcitation   = "%%CITATION = ARXIV:1809.04906;%%"
}

@article{Carta:2019hbi,
      author         = "Carta, Federico and Giacomelli, Simone and Hayashi,
                        Hirotaka and Savelli, Raffaele",
      title          = "{The Geometry of SUSY Enhancement}",
      year           = "2019",
      eprint         = "1910.09568",
      archivePrefix  = "arXiv",
      primaryClass   = "hep-th",
      reportNumber   = "DESY-19-177, ROM2F/2019/06",
      SLACcitation   = "%%CITATION = ARXIV:1910.09568;%%"
}

@article{Giacomelli:2017ckh,
      author         = "Giacomelli, Simone",
      title          = "{RG flows with supersymmetry enhancement and geometric
                        engineering}",
      journal        = "JHEP",
      volume         = "06",
      year           = "2018",
      pages          = "156",
      doi            = "10.1007/JHEP06(2018)156",
      eprint         = "1710.06469",
      archivePrefix  = "arXiv",
      primaryClass   = "hep-th",
      SLACcitation   = "%%CITATION = ARXIV:1710.06469;%%"
}

@article{Gaiotto:2009we,
      author         = "Gaiotto, Davide",
      title          = "{N=2 dualities}",
      journal        = "JHEP",
      volume         = "08",
      year           = "2012",
      pages          = "034",
      doi            = "10.1007/JHEP08(2012)034",
      eprint         = "0904.2715",
      archivePrefix  = "arXiv",
      primaryClass   = "hep-th",
      SLACcitation   = "%%CITATION = ARXIV:0904.2715;%%"
}

@article{Gaiotto:2009hg,
      author         = "Gaiotto, Davide and Moore, Gregory W. and Neitzke,
                        Andrew",
      title          = "{Wall-crossing, Hitchin Systems, and the WKB
                        Approximation}",
      year           = "2009",
      eprint         = "0907.3987",
      archivePrefix  = "arXiv",
      primaryClass   = "hep-th",
      SLACcitation   = "%%CITATION = ARXIV:0907.3987;%%"
}

@article{Maruyoshi:2018nod,
      author         = "Maruyoshi, Kazunobu and Nardoni, Emily and Song, Jaewon",
      title          = "{Landscape of Simple Superconformal Field Theories in
                        4d}",
      journal        = "Phys. Rev. Lett.",
      volume         = "122",
      year           = "2019",
      number         = "12",
      pages          = "121601",
      doi            = "10.1103/PhysRevLett.122.121601",
      eprint         = "1806.08353",
      archivePrefix  = "arXiv",
      primaryClass   = "hep-th",
      reportNumber   = "KIAS-P18044",
      SLACcitation   = "%%CITATION = ARXIV:1806.08353;%%"
}

@article{Aharony:2007dj,
      author         = "Aharony, Ofer and Tachikawa, Yuji",
      title          = "{A Holographic computation of the central charges of d=4,
                        N=2 SCFTs}",
      journal        = "JHEP",
      volume         = "01",
      year           = "2008",
      pages          = "037",
      doi            = "10.1088/1126-6708/2008/01/037",
      eprint         = "0711.4532",
      archivePrefix  = "arXiv",
      primaryClass   = "hep-th",
      reportNumber   = "WIS-20-07-NOV-DPP",
      SLACcitation   = "%%CITATION = ARXIV:0711.4532;%%"
}

@article{Romelsberger:2005eg,
	Archiveprefix = {arXiv},
	Author = {Romelsberger, Christian},
	Doi = {10.1016/j.nuclphysb.2006.03.037},
	Eprint = {hep-th/0510060},
	Journal = {Nucl. Phys.},
	Pages = {329-353},
	Primaryclass = {hep-th},
	Slaccitation = {%%CITATION = HEP-TH/0510060;%%},
	Title = {{Counting chiral primaries in N = 1, d=4 superconformal field theories}},
	Volume = {B747},
	Year = {2006}}

@article{Kinney:2005ej,
	Archiveprefix = {arXiv},
	Author = {Kinney, Justin and Maldacena, Juan Martin and Minwalla, Shiraz and Raju, Suvrat},
	Doi = {10.1007/s00220-007-0258-7},
	Eprint = {hep-th/0510251},
	Journal = {Commun. Math. Phys.},
	Pages = {209-254},
	Primaryclass = {hep-th},
	Slaccitation = {%%CITATION = HEP-TH/0510251;%%},
	Title = {{An Index for 4 dimensional super conformal theories}},
	Volume = {275},
	Year = {2007}}

@article{Xie:2012hs,
      author         = "Xie, Dan",
      title          = "{General Argyres-Douglas Theory}",
      journal        = "JHEP",
      volume         = "01",
      year           = "2013",
      pages          = "100",
      doi            = "10.1007/JHEP01(2013)100",
      eprint         = "1204.2270",
      archivePrefix  = "arXiv",
      primaryClass   = "hep-th",
      SLACcitation   = "%%CITATION = ARXIV:1204.2270;%%"
}

@article{Benvenuti:2017lle,
      author         = "Benvenuti, Sergio and Giacomelli, Simone",
      title          = "{Supersymmetric gauge theories with decoupled operators
                        and chiral ring stability}",
      journal        = "Phys. Rev. Lett.",
      volume         = "119",
      year           = "2017",
      number         = "25",
      pages          = "251601",
      doi            = "10.1103/PhysRevLett.119.251601",
      eprint         = "1706.02225",
      archivePrefix  = "arXiv",
      primaryClass   = "hep-th",
      SLACcitation   = "%%CITATION = ARXIV:1706.02225;%%"
}

@article{Maruyoshi:2016tqk,
      author         = "Maruyoshi, Kazunobu and Song, Jaewon",
      title          = "{Enhancement of Supersymmetry via Renormalization Group
                        Flow and the Superconformal Index}",
      journal        = "Phys. Rev. Lett.",
      volume         = "118",
      year           = "2017",
      number         = "15",
      pages          = "151602",
      doi            = "10.1103/PhysRevLett.118.151602",
      eprint         = "1606.05632",
      archivePrefix  = "arXiv",
      primaryClass   = "hep-th",
      reportNumber   = "IMPERIAL-TP-16-KM-02",
      SLACcitation   = "%%CITATION = ARXIV:1606.05632;%%"
}

@article{Maruyoshi:2016aim,
      author         = "Maruyoshi, Kazunobu and Song, Jaewon",
      title          = "{$ \mathcal{N}=1 $ deformations and RG flows of $
                        \mathcal{N}=2 $ SCFTs}",
      journal        = "JHEP",
      volume         = "02",
      year           = "2017",
      pages          = "075",
      doi            = "10.1007/JHEP02(2017)075",
      eprint         = "1607.04281",
      archivePrefix  = "arXiv",
      primaryClass   = "hep-th",
      reportNumber   = "IMPERIAL-TP-16-KM-03",
      SLACcitation   = "%%CITATION = ARXIV:1607.04281;%%"
}

@article{Agarwal:2016pjo,
      author         = "Agarwal, Prarit and Maruyoshi, Kazunobu and Song, Jaewon",
      title          = "{$ \mathcal{N} $ =1 Deformations and RG flows of $
                        \mathcal{N} $ =2 SCFTs, part II: non-principal
                        deformations}",
      journal        = "JHEP",
      volume         = "12",
      year           = "2016",
      pages          = "103",
      doi            = "10.1007/JHEP12(2016)103",
      eprint         = "1610.05311",
      archivePrefix  = "arXiv",
      primaryClass   = "hep-th",
      reportNumber   = "SNUTP16-006",
      SLACcitation   = "%%CITATION = ARXIV:1610.05311;%%"
}

@article{Shapere:2008zf,
      author         = "Shapere, Alfred D. and Tachikawa, Yuji",
      title          = "{Central charges of N=2 superconformal field theories in
                        four dimensions}",
      journal        = "JHEP",
      volume         = "09",
      year           = "2008",
      pages          = "109",
      doi            = "10.1088/1126-6708/2008/09/109",
      eprint         = "0804.1957",
      archivePrefix  = "arXiv",
      primaryClass   = "hep-th",
      SLACcitation   = "%%CITATION = ARXIV:0804.1957;%%"
}

@article{Argyres:1995jj,
      author         = "Argyres, Philip C. and Douglas, Michael R.",
      title          = "{New phenomena in SU(3) supersymmetric gauge theory}",
      journal        = "Nucl. Phys.",
      volume         = "B448",
      year           = "1995",
      pages          = "93-126",
      doi            = "10.1016/0550-3213(95)00281-V",
      eprint         = "hep-th/9505062",
      archivePrefix  = "arXiv",
      primaryClass   = "hep-th",
      reportNumber   = "IASSNS-HEP-95-31, RU-95-28",
      SLACcitation   = "%%CITATION = HEP-TH/9505062;%%"
}

@article{Argyres:1995xn,
      author         = "Argyres, Philip C. and Plesser, M. Ronen and Seiberg,
                        Nathan and Witten, Edward",
      title          = "{New N=2 superconformal field theories in
                        four-dimensions}",
      journal        = "Nucl. Phys.",
      volume         = "B461",
      year           = "1996",
      pages          = "71-84",
      doi            = "10.1016/0550-3213(95)00671-0",
      eprint         = "hep-th/9511154",
      archivePrefix  = "arXiv",
      primaryClass   = "hep-th",
      reportNumber   = "RU-95-81, WIS-95-59-PH, IASSNS-HEP-95-95",
      SLACcitation   = "%%CITATION = HEP-TH/9511154;%%"
}

@article{Beem:2013sza,
      author         = "Beem, Christopher and Lemos, Madalena and Liendo, Pedro
                        and Peelaers, Wolfger and Rastelli, Leonardo and van Rees,
                        Balt C.",
      title          = "{Infinite Chiral Symmetry in Four Dimensions}",
      journal        = "Commun. Math. Phys.",
      volume         = "336",
      year           = "2015",
      number         = "3",
      pages          = "1359-1433",
      doi            = "10.1007/s00220-014-2272-x",
      eprint         = "1312.5344",
      archivePrefix  = "arXiv",
      primaryClass   = "hep-th",
      reportNumber   = "YITP-SB-13-45, CERN-PH-TH-2013-311, HU-EP-13-78",
      SLACcitation   = "%%CITATION = ARXIV:1312.5344;%%"
}

@article{Cordova:2015nma,
      author         = "Cordova, Clay and Shao, Shu-Heng",
      title          = "{Schur Indices, BPS Particles, and Argyres-Douglas
                        Theories}",
      journal        = "JHEP",
      volume         = "01",
      year           = "2016",
      pages          = "040",
      doi            = "10.1007/JHEP01(2016)040",
      eprint         = "1506.00265",
      archivePrefix  = "arXiv",
      primaryClass   = "hep-th",
      SLACcitation   = "%%CITATION = ARXIV:1506.00265;%%"
}

@article{Song:2015wta,
      author         = "Song, Jaewon",
      title          = "{Superconformal indices of generalized Argyres-Douglas
                        theories from 2d TQFT}",
      journal        = "JHEP",
      volume         = "02",
      year           = "2016",
      pages          = "045",
      doi            = "10.1007/JHEP02(2016)045",
      eprint         = "1509.06730",
      archivePrefix  = "arXiv",
      primaryClass   = "hep-th",
      SLACcitation   = "%%CITATION = ARXIV:1509.06730;%%"
}

@article{Buican:2015ina,
      author         = "Buican, Matthew and Nishinaka, Takahiro",
      title          = "{On the superconformal index of Argyres-Douglas
                        theories}",
      journal        = "J. Phys.",
      volume         = "A49",
      year           = "2016",
      number         = "1",
      pages          = "015401",
      doi            = "10.1088/1751-8113/49/1/015401",
      eprint         = "1505.05884",
      archivePrefix  = "arXiv",
      primaryClass   = "hep-th",
      reportNumber   = "RU-NHETC-2015-01",
      SLACcitation   = "%%CITATION = ARXIV:1505.05884;%%"
}

@article{Kutasov:2003iy,
      author         = "Kutasov, David and Parnachev, Andrei and Sahakyan, David
                        A.",
      title          = "{Central charges and U(1)(R) symmetries in N=1
                        superYang-Mills}",
      journal        = "JHEP",
      volume         = "11",
      year           = "2003",
      pages          = "013",
      doi            = "10.1088/1126-6708/2003/11/013",
      eprint         = "hep-th/0308071",
      archivePrefix  = "arXiv",
      primaryClass   = "hep-th",
      reportNumber   = "EFI-03-40",
      SLACcitation   = "%%CITATION = HEP-TH/0308071;%%"
}

@article{Intriligator:2003jj,
      author         = "Intriligator, Kenneth A. and Wecht, Brian",
      title          = "{The Exact superconformal R symmetry maximizes a}",
      journal        = "Nucl. Phys.",
      volume         = "B667",
      year           = "2003",
      pages          = "183-200",
      doi            = "10.1016/S0550-3213(03)00459-0",
      eprint         = "hep-th/0304128",
      archivePrefix  = "arXiv",
      primaryClass   = "hep-th",
      reportNumber   = "UCSD-PTH-03-02",
      SLACcitation   = "%%CITATION = HEP-TH/0304128;%%"
}

@article{Anselmi:1997am,
      author         = "Anselmi, D. and Freedman, D. Z. and Grisaru, Marcus T.
                        and Johansen, A. A.",
      title          = "{Nonperturbative formulas for central functions of
                        supersymmetric gauge theories}",
      journal        = "Nucl. Phys.",
      volume         = "B526",
      year           = "1998",
      pages          = "543-571",
      doi            = "10.1016/S0550-3213(98)00278-8",
      eprint         = "hep-th/9708042",
      archivePrefix  = "arXiv",
      primaryClass   = "hep-th",
      reportNumber   = "BRX-TH-420, CPTH-S-553-0897, HUTP-97-A037, MIT-CTP-2666",
      SLACcitation   = "%%CITATION = HEP-TH/9708042;%%"
}

@article{Barnes:2004jj,
      author         = "Barnes, Edwin and Intriligator, Kenneth A. and Wecht,
                        Brian and Wright, Jason",
      title          = "{Evidence for the strongest version of the 4d a-theorem,
                        via a-maximization along RG flows}",
      journal        = "Nucl. Phys.",
      volume         = "B702",
      year           = "2004",
      pages          = "131-162",
      doi            = "10.1016/j.nuclphysb.2004.09.016",
      eprint         = "hep-th/0408156",
      archivePrefix  = "arXiv",
      primaryClass   = "hep-th",
      reportNumber   = "UCSD-PTH-04-09",
      SLACcitation   = "%%CITATION = HEP-TH/0408156;%%"
}

@article{Witten:1982fp,
    author = "Witten, Edward",
    editor = "Shifman, Mikhail A.",
    title = "{An SU(2) Anomaly}",
    doi = "10.1016/0370-2693(82)90728-6",
    journal = "Phys. Lett. B",
    volume = "117",
    pages = "324--328",
    year = "1982"
}

@article{Agarwal:2015vla,
    author = "Agarwal, Prarit and Intriligator, Kenneth and Song, Jaewon",
    title = "{Infinitely many $ \mathcal{N}=1 $ dualities from m + 1 \ensuremath{-} m = 1}",
    eprint = "1505.00255",
    archivePrefix = "arXiv",
    primaryClass = "hep-th",
    reportNumber = "UCSD-PTH-14-10",
    doi = "10.1007/JHEP10(2015)035",
    journal = "JHEP",
    volume = "10",
    pages = "035",
    year = "2015"
}

@article{Gadde:2015xta,
    author = "Gadde, Abhijit and Razamat, Shlomo S. and Willett, Brian",
    title = "{''Lagrangian'' for a Non-Lagrangian Field Theory with $\mathcal N=2$ Supersymmetry}",
    eprint = "1505.05834",
    archivePrefix = "arXiv",
    primaryClass = "hep-th",
    doi = "10.1103/PhysRevLett.115.171604",
    journal = "Phys. Rev. Lett.",
    volume = "115",
    number = "17",
    pages = "171604",
    year = "2015"
}

@article{Agarwal:2018ejn,
    author = "Agarwal, Prarit and Maruyoshi, Kazunobu and Song, Jaewon",
    title = "{A \textquotedblleft{}Lagrangian\textquotedblright{} for the E$_{7}$ superconformal theory}",
    eprint = "1802.05268",
    archivePrefix = "arXiv",
    primaryClass = "hep-th",
    reportNumber = "SNUTP18-001, KIAS-18001, KIAS-P18021",
    doi = "10.1007/JHEP05(2018)193",
    journal = "JHEP",
    volume = "05",
    pages = "193",
    year = "2018"
}

@article{Kang:2023dsa,
    author = "Kang, Monica Jinwoo and Lawrie, Craig and Lee, Ki-Hong and Song, Jaewon",
    title = "{Emergent N=4 Supersymmetry from N=1}",
    eprint = "2302.06622",
    archivePrefix = "arXiv",
    primaryClass = "hep-th",
    reportNumber = "CALT-TH-2023-005; DESY-23-021",
    doi = "10.1103/PhysRevLett.130.231601",
    journal = "Phys. Rev. Lett.",
    volume = "130",
    number = "23",
    pages = "231601",
    year = "2023"
}

@article{Gadde:2011uv,
    author = "Gadde, Abhijit and Rastelli, Leonardo and Razamat, Shlomo S. and Yan, Wenbin",
    title = "{Gauge Theories and Macdonald Polynomials}",
    eprint = "1110.3740",
    archivePrefix = "arXiv",
    primaryClass = "hep-th",
    reportNumber = "YITP-SB-11-30",
    doi = "10.1007/s00220-012-1607-8",
    journal = "Commun. Math. Phys.",
    volume = "319",
    pages = "147--193",
    year = "2013"
}

@article{Agarwal:2017roi,
    author = "Agarwal, Prarit and Sciarappa, Antonio and Song, Jaewon",
    title = "{$ \mathcal{N} $ =1 Lagrangians for generalized Argyres-Douglas theories}",
    eprint = "1707.04751",
    archivePrefix = "arXiv",
    primaryClass = "hep-th",
    reportNumber = "SNUTP17-003, KIAS-P17053",
    doi = "10.1007/JHEP10(2017)211",
    journal = "JHEP",
    volume = "10",
    pages = "211",
    year = "2017"
}

@article{Benvenuti:2017bpg,
    author = "Benvenuti, Sergio and Giacomelli, Simone",
    title = "{Lagrangians for generalized Argyres-Douglas theories}",
    eprint = "1707.05113",
    archivePrefix = "arXiv",
    primaryClass = "hep-th",
    reportNumber = "SISSA-32-2017-MATE-FISI",
    doi = "10.1007/JHEP10(2017)106",
    journal = "JHEP",
    volume = "10",
    pages = "106",
    year = "2017"
}

@article{Zafrir:2020epd,
    author = "Zafrir, Gabi",
    title = "{An $ \mathcal{N} $ = 1 Lagrangian for an $ \mathcal{N} $ = 3 SCFT}",
    eprint = "2007.14955",
    archivePrefix = "arXiv",
    primaryClass = "hep-th",
    doi = "10.1007/JHEP01(2021)062",
    journal = "JHEP",
    volume = "01",
    pages = "062",
    year = "2021"
}

@article{Agarwal:2014rua,
    author = "Agarwal, Prarit and Bah, Ibrahima and Maruyoshi, Kazunobu and Song, Jaewon",
    title = "{Quiver tails and $ \mathcal{N}=1 $ SCFTs from M5-branes}",
    eprint = "1409.1908",
    archivePrefix = "arXiv",
    primaryClass = "hep-th",
    reportNumber = "UCSD-PTH-14-04, CALT-TH-2014-155",
    doi = "10.1007/JHEP03(2015)049",
    journal = "JHEP",
    volume = "03",
    pages = "049",
    year = "2015"
}

@article{Razamat:2019vfd,
    author = "Razamat, Shlomo S. and Zafrir, Gabi",
    title = "{$N=1$ conformal dualities}",
    eprint = "1906.05088",
    archivePrefix = "arXiv",
    primaryClass = "hep-th",
    doi = "10.1007/JHEP09(2019)046",
    journal = "JHEP",
    volume = "09",
    pages = "046",
    year = "2019"
}

@article{Razamat:2020gcc,
    author = "Razamat, Shlomo S. and Zafrir, Gabi",
    title = "{$ \mathcal{N} $ = 1 conformal duals of gauged E$_{n}$ MN models}",
    eprint = "2003.01843",
    archivePrefix = "arXiv",
    primaryClass = "hep-th",
    doi = "10.1007/JHEP06(2020)176",
    journal = "JHEP",
    volume = "06",
    pages = "176",
    year = "2020"
}

@article{Maruyoshi:2023mnv,
    author = "Maruyoshi, Kazunobu and Nardoni, Emily and Song, Jaewon",
    title = "{Dualities of adjoint SQCD and supersymmetry enhancement}",
    eprint = "2306.08867",
    archivePrefix = "arXiv",
    primaryClass = "hep-th",
    doi = "10.1007/JHEP09(2023)082",
    journal = "JHEP",
    volume = "09",
    pages = "082",
    year = "2023"
}

@article{Zafrir:2019hps,
    author = "Zafrir, Gabi",
    title = "{An $ \mathcal{N} $ = 1 Lagrangian for the rank 1 E$_{6}$ superconformal theory}",
    eprint = "1912.09348",
    archivePrefix = "arXiv",
    primaryClass = "hep-th",
    doi = "10.1007/JHEP12(2020)098",
    journal = "JHEP",
    volume = "12",
    pages = "098",
    year = "2020"
}

@article{Cho:2024civ,
    author = "Cho, Minseok and Maruyoshi, Kazunobu and Nardoni, Emily and Song, Jaewon",
    title = "{Large landscape of 4d superconformal field theories from small gauge theories}",
    eprint = "2408.02953",
    archivePrefix = "arXiv",
    primaryClass = "hep-th",
    doi = "10.1007/JHEP11(2024)010",
    journal = "JHEP",
    volume = "11",
    pages = "010",
    year = "2024"
}

@article{Song:2017oew,
    author = "Song, Jaewon and Xie, Dan and Yan, Wenbin",
    title = "{Vertex operator algebras of Argyres-Douglas theories from M5-branes}",
    eprint = "1706.01607",
    archivePrefix = "arXiv",
    primaryClass = "hep-th",
    reportNumber = "KIAS-P17032",
    doi = "10.1007/JHEP12(2017)123",
    journal = "JHEP",
    volume = "12",
    pages = "123",
    year = "2017"
}

@article{Creutzig:2017qyf,
    author = "Creutzig, Thomas",
    title = "{W-algebras for Argyres-Douglas theories}",
    eprint = "1701.05926",
    archivePrefix = "arXiv",
    primaryClass = "hep-th",
    month = "1",
    year = "2017"
}

@article{Shapere:2008un,
    author = "Shapere, Alfred D. and Tachikawa, Yuji",
    title = "{A Counterexample to the 'a-theorem'}",
    eprint = "0809.3238",
    archivePrefix = "arXiv",
    primaryClass = "hep-th",
    doi = "10.1088/1126-6708/2008/12/020",
    journal = "JHEP",
    volume = "12",
    pages = "020",
    year = "2008"
}

@article{Aharony:1997ju,
    author = "Aharony, Ofer and Hanany, Amihay",
    title = "{Branes, superpotentials and superconformal fixed points}",
    eprint = "hep-th/9704170",
    archivePrefix = "arXiv",
    reportNumber = "RU-97-25, IASSNS-HEP-97-38",
    doi = "10.1016/S0550-3213(97)00472-0",
    journal = "Nucl. Phys. B",
    volume = "504",
    pages = "239--271",
    year = "1997"
}

@article{Aharony:1997bh,
    author = "Aharony, Ofer and Hanany, Amihay and Kol, Barak",
    title = "{Webs of (p,q) five-branes, five-dimensional field theories and grid diagrams}",
    eprint = "hep-th/9710116",
    archivePrefix = "arXiv",
    reportNumber = "IASSNS-HEP-97-113, RU-97-81, SU-ITP-97-40",
    doi = "10.1088/1126-6708/1998/01/002",
    journal = "JHEP",
    volume = "01",
    pages = "002",
    year = "1998"
}

@article{Katz:1997eq,
    author = "Katz, S. and Mayr, P. and Vafa, C.",
    title = "{Mirror symmetry and exact solution of 4-D N=2 gauge theories: 1.}",
    eprint = "hep-th/9706110",
    archivePrefix = "arXiv",
    reportNumber = "HUTP-97-A025, OSU-M-97-5, IASSNS-HEP-97-65",
    doi = "10.4310/ATMP.1997.v1.n1.a2",
    journal = "Adv. Theor. Math. Phys.",
    volume = "1",
    pages = "53--114",
    year = "1998"
}

@article{Bao:2011rc,
    author = "Bao, Ling and Pomoni, Elli and Taki, Masato and Yagi, Futoshi",
    title = "{M5-Branes, Toric Diagrams and Gauge Theory Duality}",
    eprint = "1112.5228",
    archivePrefix = "arXiv",
    primaryClass = "hep-th",
    doi = "10.1007/JHEP04(2012)105",
    journal = "JHEP",
    volume = "04",
    pages = "105",
    year = "2012"
}

@article{Chacaltana:2010ks,
    author = "Chacaltana, Oscar and Distler, Jacques",
    title = "{Tinkertoys for Gaiotto Duality}",
    eprint = "1008.5203",
    archivePrefix = "arXiv",
    primaryClass = "hep-th",
    reportNumber = "UTTG-11-10, TCC-020-10",
    doi = "10.1007/JHEP11(2010)099",
    journal = "JHEP",
    volume = "11",
    pages = "099",
    year = "2010"
}

@article{Bergman:2014kza,
    author = "Bergman, Oren and Zafrir, Gabi",
    title = "{Lifting 4d dualities to 5d}",
    eprint = "1410.2806",
    archivePrefix = "arXiv",
    primaryClass = "hep-th",
    doi = "10.1007/JHEP04(2015)141",
    journal = "JHEP",
    volume = "04",
    pages = "141",
    year = "2015"
}

@article{Nanopoulos:2010bv,
    author = "Nanopoulos, Dimitri and Xie, Dan",
    title = "{More Three Dimensional Mirror Pairs}",
    eprint = "1011.1911",
    archivePrefix = "arXiv",
    primaryClass = "hep-th",
    reportNumber = "MIFPA-10-51",
    doi = "10.1007/JHEP05(2011)071",
    journal = "JHEP",
    volume = "05",
    pages = "071",
    year = "2011"
}

@article{Benvenuti:2017kud,
    author = "Benvenuti, Sergio and Giacomelli, Simone",
    title = "{Abelianization and sequential confinement in $2+1$ dimensions}",
    eprint = "1706.04949",
    archivePrefix = "arXiv",
    primaryClass = "hep-th",
    doi = "10.1007/JHEP10(2017)173",
    journal = "JHEP",
    volume = "10",
    pages = "173",
    year = "2017"
}

@article{Buican:2015hsa,
    author = "Buican, Matthew and Nishinaka, Takahiro",
    title = "{Argyres{\textendash}Douglas theories, S$^1$ reductions, and topological symmetries}",
    eprint = "1505.06205",
    archivePrefix = "arXiv",
    primaryClass = "hep-th",
    reportNumber = "RU-NHETC-2015-02",
    doi = "10.1088/1751-8113/49/4/045401",
    journal = "J. Phys. A",
    volume = "49",
    number = "4",
    pages = "045401",
    year = "2016"
}

@article{Dedushenko:2019mnd,
    author = "Dedushenko, Mykola and Wang, Yifan",
    title = "{4d/2d {\textrightarrow} 3d/1d: A song of protected operator algebras}",
    eprint = "1912.01006",
    archivePrefix = "arXiv",
    primaryClass = "hep-th",
    reportNumber = "CALT-TH 2019-041, PUPT-2602",
    doi = "10.4310/ATMP.2022.v26.n7.a2",
    journal = "Adv. Theor. Math. Phys.",
    volume = "26",
    number = "7",
    pages = "2011--2075",
    year = "2022"
}

@article{Benvenuti:2018bav,
    author = "Benvenuti, Sergio",
    title = "{A tale of exceptional $3d$ dualities}",
    eprint = "1809.03925",
    archivePrefix = "arXiv",
    primaryClass = "hep-th",
    doi = "10.1007/JHEP03(2019)125",
    journal = "JHEP",
    volume = "03",
    pages = "125",
    year = "2019"
}

@article{Gang:2018huc,
    author = "Gang, Dongmin and Yamazaki, Masahito",
    title = "{Three-dimensional gauge theories with supersymmetry enhancement}",
    eprint = "1806.07714",
    archivePrefix = "arXiv",
    primaryClass = "hep-th",
    reportNumber = "IPMU18-0081",
    doi = "10.1103/PhysRevD.98.121701",
    journal = "Phys. Rev. D",
    volume = "98",
    number = "12",
    pages = "121701",
    year = "2018"
}

@article{Gang:2023rei,
    author = "Gang, Dongmin and Kim, Heeyeon and Stubbs, Spencer",
    title = "{Three-Dimensional Topological Field Theories and Nonunitary Minimal Models}",
    eprint = "2310.09080",
    archivePrefix = "arXiv",
    primaryClass = "hep-th",
    doi = "10.1103/PhysRevLett.132.131601",
    journal = "Phys. Rev. Lett.",
    volume = "132",
    number = "13",
    pages = "131601",
    year = "2024"
}

@article{Xie:2016evu,
    author = "Xie, Dan and Yan, Wenbin and Yau, Shing-Tung",
    title = "{Chiral algebra of the Argyres-Douglas theory from M5 branes}",
    eprint = "1604.02155",
    archivePrefix = "arXiv",
    primaryClass = "hep-th",
    doi = "10.1103/PhysRevD.103.065003",
    journal = "Phys. Rev. D",
    volume = "103",
    number = "6",
    pages = "065003",
    year = "2021"
}

@article{Hwang:2021xyw,
    author = "Hwang, Chiung and Razamat, Shlomo S. and Sabag, Evyatar and Sacchi, Matteo",
    title = "{Rank $Q$ E-string on spheres with flux}",
    eprint = "2103.09149",
    archivePrefix = "arXiv",
    primaryClass = "hep-th",
    doi = "10.21468/SciPostPhys.11.2.044",
    journal = "SciPost Phys.",
    volume = "11",
    number = "2",
    pages = "044",
    year = "2021"
}

@article{Gaiotto:2024ioj,
    author = "Gaiotto, Davide and Kim, Heeyeon",
    title = "{3D TFTs from 4d $ \mathcal{N} $ = 2 BPS particles}",
    eprint = "2409.20393",
    archivePrefix = "arXiv",
    primaryClass = "hep-th",
    doi = "10.1007/JHEP03(2025)173",
    journal = "JHEP",
    volume = "03",
    pages = "173",
    year = "2025"
}

@article{ArabiArdehali:2024vli,
    author = "Arabi Ardehali, Arash and Gang, Dongmin and Rajappa, Neville Joshua and Sacchi, Matteo",
    title = "{3d SUSY enhancement and non-semisimple TQFTs from four dimensions}",
    eprint = "2411.00766",
    archivePrefix = "arXiv",
    primaryClass = "hep-th",
    reportNumber = "YITP-SB-2024-26",
    doi = "10.1007/JHEP09(2025)179",
    journal = "JHEP",
    volume = "09",
    pages = "179",
    year = "2025"
}

@article{ArabiArdehali:2024ysy,
    author = "Arabi Ardehali, Arash and Dedushenko, Mykola and Gang, Dongmin and Litvinov, Mikhail",
    title = "{Bridging 4D QFTs and 2D VOAs via 3D high-temperature EFTs}",
    eprint = "2409.18130",
    archivePrefix = "arXiv",
    primaryClass = "hep-th",
    reportNumber = "YITP-SB-2024-13",
    doi = "10.1007/JHEP02(2026)038",
    journal = "JHEP",
    volume = "02",
    pages = "038",
    year = "2026"
}

@article{Dedushenko:2023cvd,
    author = "Dedushenko, Mykola",
    title = "{On the 4d/3d/2d view of the SCFT/VOA correspondence}",
    eprint = "2312.17747",
    archivePrefix = "arXiv",
    primaryClass = "hep-th",
    month = "12",
    year = "2023"
}

@article{Rastelli:2023sfk,
    author = "Rastelli, Leonardo and Rayhaun, Brandon C.",
    title = "{Rationality in four dimensions}",
    eprint = "2308.06312",
    archivePrefix = "arXiv",
    primaryClass = "hep-th",
    doi = "10.1103/PhysRevD.109.105018",
    journal = "Phys. Rev. D",
    volume = "109",
    number = "10",
    pages = "105018",
    year = "2024"
}

@article{Dolan:2008qi,
    author = "Dolan, F. A. and Osborn, H.",
    title = "{Applications of the Superconformal Index for Protected Operators and q-Hypergeometric Identities to N=1 Dual Theories}",
    eprint = "0801.4947",
    archivePrefix = "arXiv",
    primaryClass = "hep-th",
    reportNumber = "DAMTP-08-07, DIAS-STP-08-02, SHEP-08-06",
    doi = "10.1016/j.nuclphysb.2009.01.028",
    journal = "Nucl. Phys. B",
    volume = "818",
    pages = "137--178",
    year = "2009"
}

@article{Kim:2025klh,
    author = "Kim, Heeyeon and Kim, Hongseok and Song, Jaewon",
    title = "{Macdonald index from 3d TQFT}",
    eprint = "2511.11186",
    archivePrefix = "arXiv",
    primaryClass = "hep-th",
    doi = "10.1007/JHEP03(2026)213",
    journal = "JHEP",
    volume = "03",
    pages = "213",
    year = "2026"
}

\end{document}